\documentclass[oupdraft]{bio}
\usepackage{url} % not crucial - just used below for the URL 
\usepackage{algorithm}
\usepackage{algorithmic}
\usepackage{mathrsfs}
\usepackage[margin=1in]{geometry}
\usepackage{graphicx}
\usepackage{amsmath, amssymb}
\usepackage{bbold}
\usepackage[colorlinks=true,allcolors=blue]{hyperref}
\usepackage{mathbbol}
\usepackage{float}
\usepackage{xcolor}
\usepackage{bm} % Math packages
\usepackage{setspace}
\usepackage{multirow}
\usepackage{lscape}
\usepackage[figuresright]{rotating}
\usepackage{adjustbox}
\usepackage{booktabs}
\usepackage{ragged2e}
\usepackage{subcaption}
\usepackage{float}
\usepackage{amsthm}
\usepackage{mathbbol}
\usepackage{tikz}
\usepackage{fix-cm}

\newcommand{\bpm}{\begin{pmatrix}}
\newcommand{\epm}{\end{pmatrix}}

\theoremstyle{definition}

\def\*#1{\mathbf{#1}}
\def\^#1{\amsmathbb{#1}}
\def\##1{\mathbb{#1}}
\DeclareSymbolFontAlphabet{\amsmathbb}{AMSb}%

\begin{document}
% Title of paper
\title{Variable Selection in Functional Linear Quantile Regression  for Identifying Associations between Daily Patterns of Physical Activity and Cognitive Function}

% List of authors, with corresponding author marked by asterisk
\author{Yuanzhen Yue$^{1}$, Stella Self$^{1}$, Yichao Wu$^{2}$, Jiajia Zhang$^{1}$, and Rahul Ghosal$^{1,\ast}$ \\[4pt]
% Author addresses
\textit{$^{1}$ Department of Epidemiology and Biostatistics, University of South Carolina \\
$^{2}$ Department of Mathematics, Statistics, and Computer Science, University of Illinois at Chicago}
\\[2pt]
% E-mail address for correspondence
{rghosal@mailbox.sc.edu}}

% Running headers of paper:
\markboth%
% First field is the short list of authors
{Y. Yue and others}
% Second field is the short title of the paper
{VSFLQR}

\maketitle
% Add a footnote for the corresponding author if one has been
% identified in the author list
\footnotetext{To whom correspondence should be addressed.}

\begin{abstract}
{Quantile regression is useful for characterizing the conditional distribution of a response variable and understanding heterogeneity in the covariate effects at different quantiles. The rise of high-dimensional physiological data in biomedical research through wearable and sensor devices underscores the need for effective variable selection methods for interpretable and accurate quantile regression, which can offer robust insights into heterogeneous and dynamic covariate effects. We develop a flexible variable selection approach for functional linear quantile regression with multiple functional and scalar predictors. We use a smooth approximation of the quantile loss function and integrate functional principal component analysis (FPCA) with a group minimax concave penalty (MCP) to impose sparsity on the functional coefficients. A computationally efficient group descent algorithm is employed for optimization. Through numerical simulations, we demonstrate a satisfactory selection, estimation, and prediction accuracy of the proposed method across different quantiles for both dense and sparsely observed functional data. The proposed method is applied to accelerometer data from the 2011–2014 National Health and Nutrition Examination Survey (NHANES) to identify key time-varying distributional patterns of physical activity and demographic predictors associated with cognitive function across different quantiles. Our analysis provides new insights into the complex relationship between the daily distributional patterns of physical activity and cognitive function among older adults, capturing heterogeneous associations across different quantiles.}
{Functional Linear Quantile Regression; Variable Selection; NHANES; Cognitive Function; Physical Activity.}
\end{abstract}

\section{Introduction}
\label{intro}
Quantile regression (QR), first introduced by \cite{koenker1978regression}, has become a widely used statistical tool for modeling conditional quantiles of a response variable. Unlike standard regression approaches that focus on modeling the conditional mean, QR provides a more comprehensive view of the response distribution, making it especially useful in applications where tail behavior or heteroskedasticity is of interest. The key advantage of QR lies in its ability to characterize the entire conditional distribution of a response variable, rather than summarizing it through a single central tendency measure. This property makes QR particularly useful for analyzing asymmetric or heavy-tailed distributions, which are common in economic, biomedical, and environmental studies \citep{davino2013quantile,koenker2017quantile}. Moreover, QR is inherently robust to outliers in the response variable because it minimizes an asymmetric loss function, reducing the influence of extreme values compared to ordinary least squares (OLS) regression \citep{koenker2001quantile,koenker2005quantile}. These features have led to the widespread adoption of QR in disciplines such as Bayesian \citep{yu2001bayesian} and machine learning applications \citep{takeuchi2006nonparametric}.

Functional data analysis (FDA) \citep{Ramsay05functionaldata, crainiceanu2024functional} is a statistical framework for analyzing data that are naturally represented as functions, curves, or trajectories observed over a continuum such as time, space, or other domains. In contrast to traditional multivariate methods, which treat observations as finite-dimensional vectors, FDA focuses on infinite-dimensional objects, making it particularly well suited for complex data structures in which observations are more appropriately viewed as smooth curves or surfaces. FDA has a wide range of applications across diverse scientific disciplines. In the biomedical sciences, it has been applied to the analysis of growth curves \citep{ramsay2007applied}, heart rate signals \citep{ratcliffe2002functional, diller2006heart}, physical activity (PA) profiles \citep{xiao2015quantifying, goldsmith2016new, cui2021additive, cui2022fast, ghosal2023functional}, and brain activity patterns \citep{tian2010functional}, providing deeper insights into underlying physiological processes and health outcomes. In environmental studies, FDA facilitates the analysis of temporal and spatial data, such as climate dynamics and pollution levels, enabling flexible modeling and improved prediction of environmental phenomena \citep{besse2000autoregressive}.

The integration of functional data analysis (FDA) with quantile regression is particularly valuable in biomedical research, where continuous monitoring of physiological signals is increasingly common. Functional measurements capture rich temporal dynamics that may carry important prognostic information for the conditional distribution of health outcomes of interest. By leveraging the FDA, these functional covariates can be modeled and analyzed in a principled manner, leading to a more comprehensive understanding of their associations with outcomes. Functional quantile regression (FQR) \citep{kato2012estimation} extends the classical QR by incorporating functional predictors. A variety of methods have been developed for estimation and inference in FQR \citep{tang2014partial,yao2017regularized,du2018estimation,li2022inference,wang2023smoothed} and its extensions.

Advances in technology have led to a dramatic increase in the complexity of biomedical data, particularly in contemporary studies that involve high-dimensional physiological signals. Such studies routinely generate large volumes of data, creating substantial challenges in identifying the truly relevant predictors associated with the outcome of interest. Robust variable selection methods are therefore essential for enhancing both interpretability and predictive accuracy in high-dimensional models \citep{hastie2015statistical}. A wide range of methods has been developed for variable selection in classical quantile regression with scalar covariates, primarily through penalized and Bayesian approaches. These include smoothly clipped absolute deviation (SCAD) and adaptive-LASSO \citep{wu2009variable}, the elastic net \citep{su2021elastic}, and variational Bayesian method \citep{dai2023high}.

In this article, our motivating application is based on triaxial accelerometer and cognitive functioning data collected during the 2011-2014 waves of the National Health and Nutrition Examination Survey (NHANES).  In NHANES 2011-2014, minute-level accelerometry data are reported in Monitor Independent Movement Summary (MIMS) units \citep{john2019open}, an open source device-independent metric for summarizing PA. The animal fluency test (CFDAST) was administered to participants aged 60 years and older in NHANES 2011-2014, and serves as our cognitive outcome of interest. The animal fluency test (CFDAST) score examines categorical verbal fluency, a component of executive function, and has been shown previously to discriminate between persons with normal cognitive functioning and those with mild cognitive impairment, or Alzheimer’s disease \citep{henry2004verbal,canning2004diagnostic}.

\begin{figure}[H]
  \centering
    \begin{subfigure}{\textwidth}
        \centering
        \includegraphics[scale=0.35]{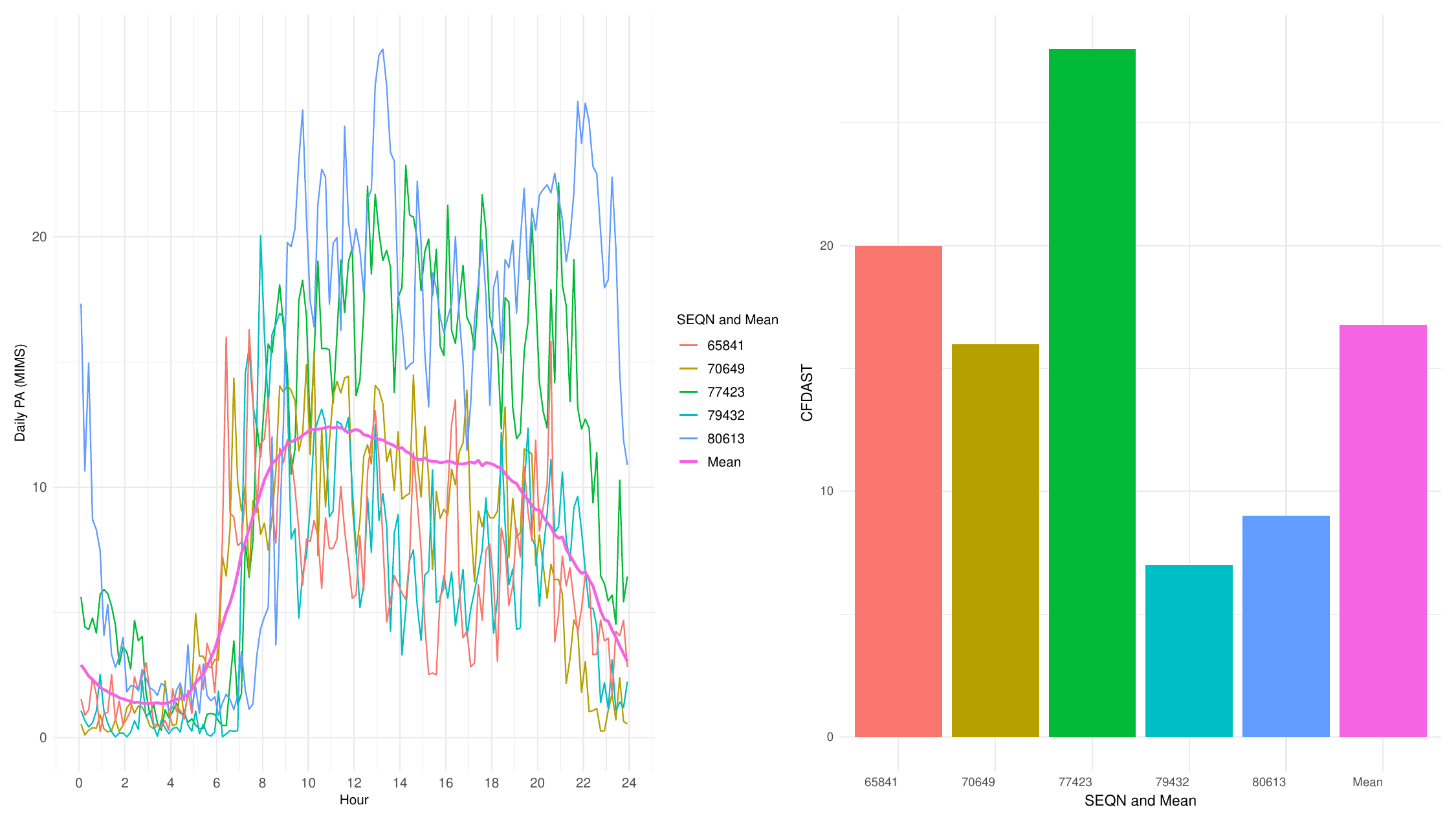}
        \caption{The average diurnal PA (MIMS) across all participants, along with the diurnal PA and CFDAST score for five randomly selected participants by SEQN (unique subject identifier).}
        \label{2a}
    \end{subfigure}
    
    \begin{subfigure}{\textwidth}
        \centering
        \includegraphics[scale=0.35]{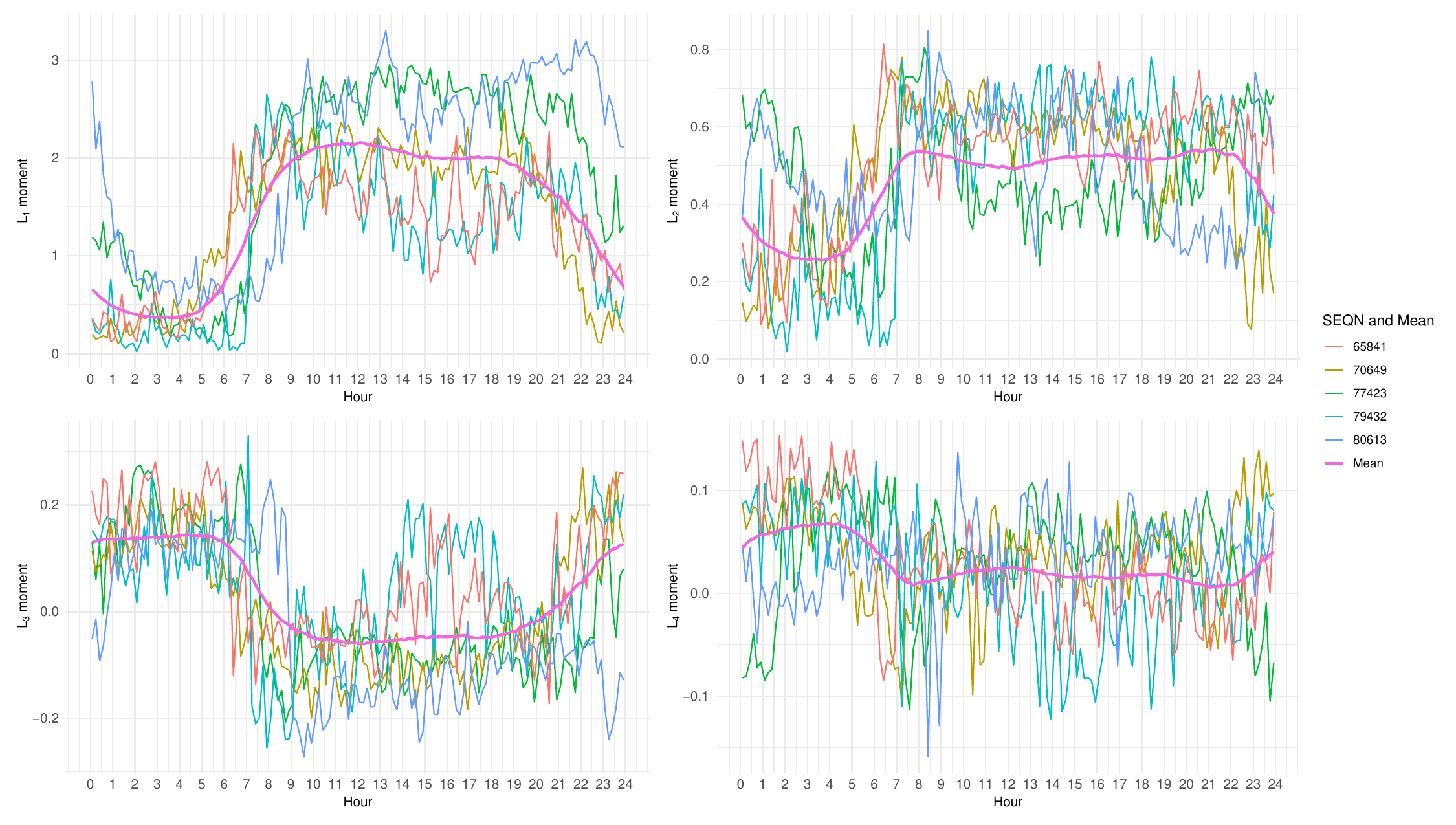}
        \caption{The average of the first four time-varying diurnal L-moments (capturing daily distributional patterns of physical activity) across all participants, as well as the first four L-moments for five randomly selected participants by SEQN (unique subject identifier), using log-transformed MIMS.}
        \label{2b}
    \end{subfigure}
\caption{Visual summary of physical activity across participants.}
\label{fig1}
\end{figure}

Cognitive scores are useful for examining the association of cognitive functioning with other medical conditions and risk factors \citep{zhou2024dietary}, and for tracking cognitive decline in the aging population \citep{anderson2007cognitive}. Most existing analyses examining the association between physical activity and cognitive functioning rely on summary-level PA metrics \citep{campbell2023estimating, quinlan2023physical, yau2025physical}. More recently, functional regression methods have been used to examine how the daily mean activity patterns relate to the average cognitive function \citep{ghosal2022scalar} in the population. 
Figure~\ref{fig1} (Panel a) presents the average diurnal PA (in MIMS) across all participants, together with activity trajectories and CFDAST score for five randomly selected individuals. Emerging work in distributional analysis \citep{ghosal2022scalar,cho2024exploring,yue2026variable} also suggests that daily PA patterns beyond the mean, such as variability and higher-order characteristics that evolve over the course of the day, may offer additional insight beyond average activity patterns. The diurnal L-moments \citep{ghosal2022scalar, cho2024exploring,yue2026variable} provide a convenient way to summarize these time-varying daily distributional patterns of PA. Figure~\ref{fig1} (Panel b) shows the population averages of the first four L-moments, along with the corresponding curves for five randomly chosen participants, computed from log-transformed activity counts. The primary research question in this article is to understand how the conditional distribution of the cognitive function depends on these daily physical activity patterns beyond the daily mean activity and identify the influential daily PA patterns together with the key demographic and lifestyle variables.

Despite the growing availability of high-dimensional functional data such as physical activity, heart rate, and energy expenditure, there are few works that address variable selection in quantile regression that simultaneously accommodates both functional and scalar predictors. Previously \cite{ma2019quantile} proposed a high-dimensional partially linear functional quantile regression that employs the group SCAD penalty to identify important scalar predictors and functional predictors. Their approach relies on a two-step tuning procedure and direct optimization of the non-smooth quantile loss. Compared to this method, our approach uses a smoothed quantile loss and a different sparsity penalty, resulting in a computationally efficient algorithm and providing smooth and sparse estimates of functional coefficients in the functional linear quantile regression. In this paper, we propose a variable selection method for a functional linear quantile regression (FLQR) that accommodates multiple functional and scalar covariates. Our contributions are threefold. First, we show that variable selection in a functional linear quantile regression with both functional and scalar predictors can be naturally formulated as a group selection problem. Second, we employ a functional principal component analysis (FPCA) based representation of the functional covariates naturally applicable to dense or sparse functional data, together with a group minimax concave penalty (MCP), to jointly encourage data-adaptive smoothness and sparsity in the functional effects, while also facilitating selection of important scalar predictors. Third, we employ an efficient group descent algorithm for model fitting and propose an automated extended Bayesian information criterion (EBIC) for selecting the tuning parameters.

The remainder of this paper is organized as follows. Section~\ref{sec:method1} introduces the proposed modeling framework and describes the variable selection procedure. Section~\ref{sec:sim_stud} presents simulation studies evaluating the performance of the proposed method. In Section~\ref{realdat}, we apply our approach to the NHANES 2011-2014 accelerometer data and present our research findings. Finally, Section~\ref{disc} concludes with a discussion and outlines potential directions for future research.

\section{Methodology}
\label{sec:method1}
\subsection{Modeling Framework}
\label{mf1}
For each subject \( i \), we observe a response variable \( Y_i \) along with a set of scalar covariates represented by the vector \( \boldsymbol{X}_i = (X_{i0}, X_{i1}, \dots, X_{iB})^{\top} \), which includes an intercept given by $X_{i0}$. Additionally, we consider multiple functional covariates, denoted as \( Z_{ij}(s) \in \mathscr{L}^2 [0, 1] \), for \( j = 1, 2, \dots, J \). We assume that the functional covariates lie in a real, separable Hilbert space, which we take to be $\mathscr{L}^2[0,1]$ throughout this paper. We further assume that these functional covariates are observed on a dense, regular grid $S = \{t_1, t_2, \ldots, t_m\} \subset [0,1]$, consistent with our motivating application, and for notational simplicity. This assumption, however, can be relaxed, and the proposed method can be directly applied to more general settings, such as sparsely observed functional data \citep{yao2005functional}. Let the data observed for the $i$th subject be denoted by
$D_i = \{Y_i, \boldsymbol{X}_i, Z_{ij}(s)\}, i = 1,2,\ldots,n$. We assume that observations from different subjects are mutually independent.

\subsection{Functional Linear Quantile Regression}
\begin{figure}[H]
    \centering
    \includegraphics[scale=0.5]{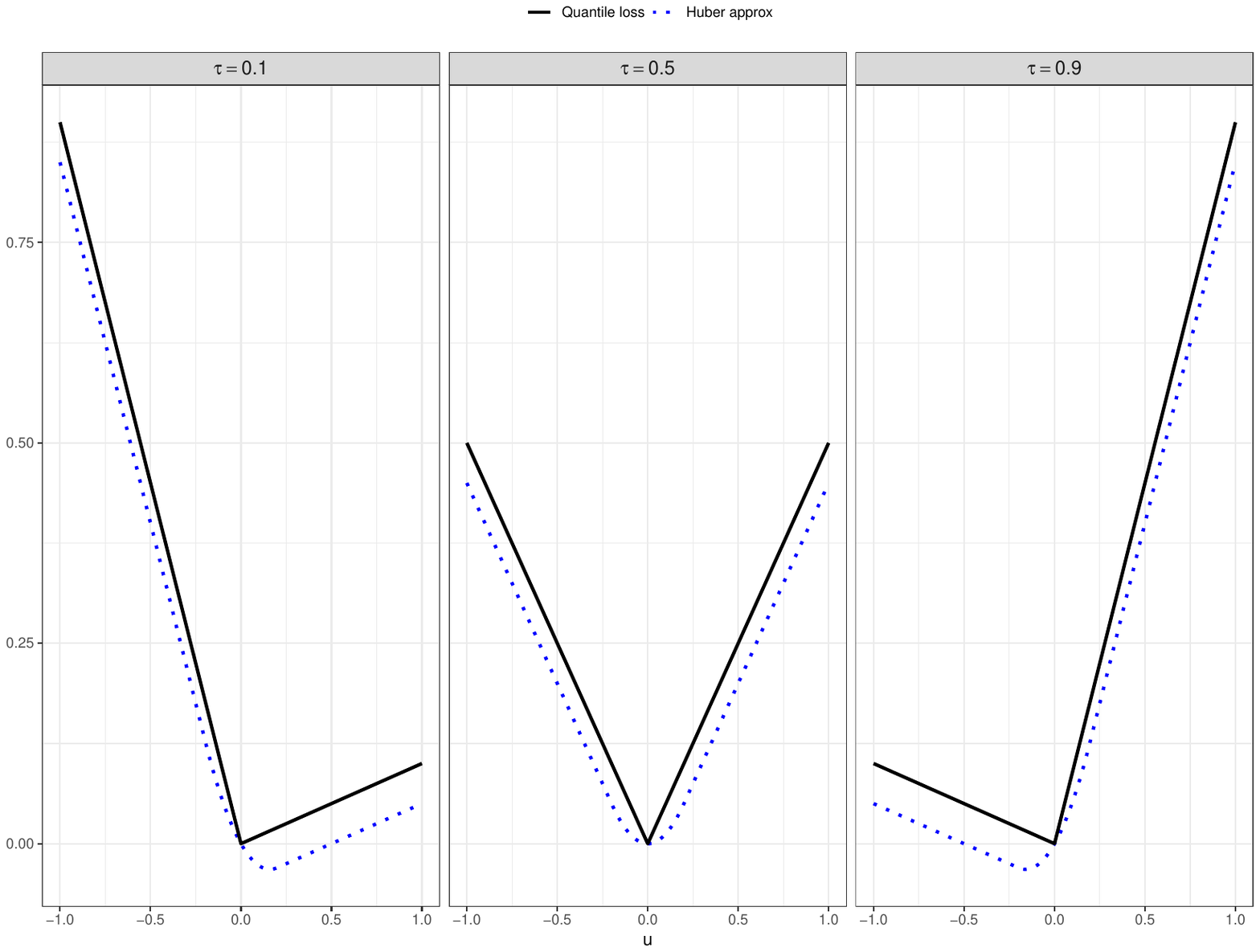}
    \caption{Quantile loss vs. Huber-approximated quantile loss.}
    \label{fig2}
\end{figure}

We consider a functional linear quantile regression (FLQR) model \citep{kato2012estimation} given by:
\begin{equation}\label{eq2_2}
    Q_{Y_i} (\tau | \boldsymbol{X}_i, Z_{i1}(\cdot), \dots, Z_{iJ}(\cdot)) = \boldsymbol{X}_i^{\top} \boldsymbol{\beta} (\tau) + \sum_{j=1}^{J} \int_0^1 Z_{ij}(s) \Gamma_j(s, \tau) ds,
\end{equation}
where \( \boldsymbol{\beta}(\tau) = (\beta_1(\tau),\dots,\beta_B(\tau))^{\top} \) is the vector of scalar coefficients dependent on the quantile \(\tau\), and \( \Gamma_j(\cdot,\tau) \) denotes the functional coefficient corresponding to the \( j \)-th predictor, varying smoothly across both quantile levels and the functional domain. For notational convenience, we set \(\boldsymbol{\Gamma}(\cdot,\tau)=(\Gamma_1(\cdot,\tau),\dots,\Gamma_J(\cdot,\tau))\), and summarize the linear predictor for subject \(i\) as: \(\eta_i = \boldsymbol{X}_i^{\top}\boldsymbol{\beta}(\tau)+\sum_{j=1}^{J}\int_0^1 Z_{ij}(s)\Gamma_j(s,\tau)\,ds\). We assume that the covariates $Z_{ij}(s)$ have been centered, without loss of generality. We are interested in selecting the important scalar and functional predictors in FLQR, which amounts to imposing sparsity in $\boldsymbol{\beta}(\tau),\boldsymbol{\Gamma}(\cdot,\tau)$. For Traditional FLQR (without sparsity), estimating the regression coefficients \((\boldsymbol{\beta}(\tau), \boldsymbol{\Gamma}(\cdot,\tau))\) for a given quantile level \(\tau \in (0,1)\) involves solving the following optimization problem:
\begin{equation}
    \min_{\boldsymbol{\beta}(\tau), \boldsymbol{\Gamma}(\cdot,\tau)} \sum_{i=1}^{n} \rho_{\tau} \left( Y_i - \boldsymbol{X}_i^{\top} \boldsymbol{\beta} (\tau) + \sum_{j=1}^{J} \int_0^1 Z_{ij}(s) \Gamma_j(s, \tau)ds \right),
\end{equation}
where \(\rho_\tau(u)\) is the quantile loss function, commonly referred to as the check function, introduced by \cite{koenker1978regression}. It is defined as: $\rho_\tau(u) = u[\tau - I(u < 0)]$. Alternatively, the quantile loss can be equivalently expressed as: $\rho_\tau(u) = \frac{1}{2} [|u| + (2\tau - 1)u]$. While the check function provides an effective framework for quantile estimation, its non-differentiability poses significant computational challenges, particularly for large datasets and high-dimensional functional predictors, and when combined with nonconvex penalties. To address this issue, a smooth approximation is often preferred. One widely used approach is the Huber loss function, originally proposed by \cite{huber1992robust}, which approximates the absolute value function \(|u|\) in a computationally efficient manner:
\begin{equation}
   h_\gamma(u) = 
\begin{cases}
\frac{u^2}{2\gamma}, & \text{if } |u| \leq \gamma, \\
|u| - \frac{\gamma}{2}, & \text{if } |u| > \gamma.
\end{cases} 
\end{equation}
This formulation smoothly transitions from a squared loss for small residuals to an absolute loss for larger deviations, thereby reducing sensitivity to outliers while maintaining differentiability. Leveraging this approximation, we will employ the Huber-based quantile loss function \citep{yi2017semismooth}: $h_\gamma^\tau(u) = h_\gamma(u) + (2\tau - 1)u$ for defining our penalized objective function. This formulation ensures that the advantages of the standard quantile loss function are retained while introducing computational benefits through the Huber approximation. Figure~\ref{fig2} shows how the original quantile loss $\rho_\tau(u)$ can be approximated by Huber-approximated quantile loss for $\tau \in \{0.1, 0.5, 0.9\}$ with $\gamma=0.2$. By using a smooth approximation of the loss function, optimization algorithms converge more efficiently, particularly in large-scale or high-dimensional settings.
Estimating both \(\Gamma_j(\cdot, \tau)\) and \(\boldsymbol{\beta}(\tau)\) simultaneously presents a significant challenge due to the infinite-dimensional nature of \(\Gamma_j(\cdot, \tau)\). Directly treating the functional parameter as a high-dimensional multivariate counterpart leads to an excessively large parameter space, making both theoretical estimation with desirable convergence properties and practical computation infeasible. To circumvent this curse of dimensionality, we leverage functional principal component analysis (FPCA) technique \citep{yao2005functional}. This method allows the functional covariates to be effectively represented in a finite-dimensional space by projecting them onto a set of data-driven basis functions. To characterize the covariance structure of \( Z_{ij}(\cdot) \), we define its covariance function as 
\(G_j(s,t) = \text{Cov}(Z_{j}(s), Z_{j}(t))\), which captures the dependency between different time points \( s, t \in [0,1] \). Utilizing the Mercer’s Theorem, we express \( G_j(s,t) \) in terms of its eigenfunctions \(\{\phi_{jq}\}_{q=1,2,\dots}\) and associated non-increasing eigenvalues \(\{\lambda_{jq}\}_{q=1,2,\dots}\) as follows: \(G_j(s,t) = \sum_{q} \lambda_{jq} \phi_{jq}(s) \phi_{jq}(t)\). According to the Karhunen-Loève expansion, each \( Z_{ij}(s) \) (centered) can be represented as a linear combination of orthonormal basis functions: \(Z_{ij}(s) = \sum_{q=1}^{Q_j} \xi_{ijq} \phi_{jq}(s)\), where the number of eigenfunctions, \( Q_j \), is determined using the percentage of variance explained (PVE) criterion. 
%Throughout this paper, we adopt a PVE threshold of $95\%$.%
The corresponding principal component scores are computed as: \(\xi_{ijq} = \int_0^1 Z_{ij}(s) \phi_{jq}(s) ds\). These scores are uncorrelated random variables with zero mean and variance \( E\xi_{ijq}^2 = \lambda_{jq} \), provided that \( \sum_q \lambda_{jq} < \infty \). By substituting this expansion into the model \ref{eq2_2}, we can reformulate it as:
\begin{equation}
    \begin{split}
    \mathnormal{Q}_{Y_i} (\tau|\boldsymbol{X}_i, Z_{i1}(\cdot), \dots, Z_{iJ}(\cdot)) &= \boldsymbol{X}_i^{\top} \boldsymbol{\beta} (\tau) + \sum_{j=1}^{J} \int_0^1 Z_{ij}(s) \Gamma_{j}(s,\tau) ds \\
    &= \boldsymbol{X}_i^{\top} \boldsymbol{\beta} (\tau) + \sum_{j=1}^{J} \int_0^1 \left( \sum_{q=1}^{Q_j} \xi_{ijq} \phi_{jq}(s) \right) \Gamma_{j}(s,\tau) ds \\
    &= \boldsymbol{X}_i^{\top} \boldsymbol{\beta} (\tau) + \sum_{j=1}^{J} \sum_{q=1}^{Q_j} \xi_{ijq} \int_0^1 \phi_{jq}(s) \Gamma_{j}(s,\tau) ds \\
    &= \boldsymbol{X}_i^{\top} \boldsymbol{\beta} (\tau) + \sum_{j=1}^{J} \sum_{q=1}^{Q_j} \xi_{ijq} \alpha_{jq} (\tau) \\
    &= \boldsymbol{X}_i^{\top} \boldsymbol{\beta} (\tau) + \sum_{j=1}^{J} \boldsymbol{\xi}_{ij}^{\top} \boldsymbol{\alpha_j}(\tau),
\end{split}
\end{equation}
where $\alpha_{jq} (\tau)=\int_0^1 \phi_{jq}(s) \Gamma_{j}(s,\tau) ds$, $\boldsymbol{\xi}_i=(\boldsymbol{\xi}_{i1}^{\top},\ldots,\boldsymbol{\xi}_{iJ}^{\top})^{\top}$, $\boldsymbol{\alpha_j} (\tau) = (\alpha_{j_1}, \alpha_{j_2}, \dots, \alpha_{j_{Q_j}})^{\top}$, and $\boldsymbol{\alpha} (\tau) = (\boldsymbol{\alpha_1} (\tau)^{\top}, \boldsymbol{\alpha_2} (\tau)^{\top}, \dots, \boldsymbol{\alpha_J} (\tau)^{\top})^{\top}$. In order to achieve sparsity in $\boldsymbol{\beta}(\tau),\boldsymbol{\Gamma}(\cdot,\tau)$, we recognize the variable selection problem as performing a group selection, where covariates $ \boldsymbol{\xi}_{ij}^{\top}$ naturally define the groupings. For example, one can estimate \(\boldsymbol{\beta}(\tau)\) and \(\boldsymbol{\alpha}(\tau)\) by minimizing a penalized Huber-based quantile loss function with group LASSO penalty \citep{yuan2006model} on the scores to induce sparsity on the coefficients, and a LASSO penalty on scalar coefficients,
\begin{equation}\label{eqlasso}
\resizebox{0.9\textwidth}{!}{$
\displaystyle
\begin{split}
(\hat{\boldsymbol{\beta}}(\tau), \hat{\boldsymbol{\alpha}}(\tau))
&= \underset{\boldsymbol{\beta} (\tau), \boldsymbol{\alpha} (\tau)}{\arg \min}
\left\{ \frac{1}{2n}
\Big( h_\gamma^\tau(Y_i -\eta_i)
+ \sum_{b=1}^{B} \lambda \|\beta_b (\tau)\|_1
+ \sum_{j=1}^{J} \lambda_j \|\boldsymbol{\alpha}_j (\tau)\|_2 \Big) \right\} \\
&= \underset{\boldsymbol{\beta} (\tau), \boldsymbol{\alpha} (\tau)}{\arg \min}
\left\{ \frac{1}{2n}
\Big( h_\gamma^\tau(Y_i -\eta_i)
+ \sum_{b=1}^{B} P_{\text{LASSO}, \lambda,\phi}(\|\beta_b (\tau)\|_1)
+ \sum_{j=1}^{J} P_{\text{LASSO}, \lambda_j,\phi}(\|\boldsymbol{\alpha}_j (\tau)\|_2) \Big) \right\}.
\end{split}
$}
\end{equation}

\subsection{VSFLQR Method for Variable Selection in FLQR}
In this paper, we propose the VSFLQR method for performing variable selection in FLQR, using a group MCP \citep{zhang2010nearly} extension of the above optimization criterion \ref{eqlasso}. Although LASSO \citep{tibshirani1996regression} is widely used for variable selection in high-dimensional settings, it is well known to suffer from a relatively high false positive rate and to produce biased coefficient estimates \citep{mazumder2011sparsenet}. The group MCP \citep{zhang2010nearly} is a nonconvex penalty designed to alleviate this bias by gradually relaxing the penalization on coefficients as their magnitudes increase \citep{breheny2015group}. As a result, MCP strikes an effective balance between sparsity and unbiased estimation, making it particularly well suited for variable selection in functional regression models \citep{chen2016variable, ghosal2020variable, ghosal2023variable,yue2026variable}. Moreover, MCP enjoys several desirable theoretical properties, including the oracle property under standard regularity conditions \citep{zhang2010nearly, breheny2015group}. These advantages motivate its use in our proposed VSFLQR method, where MCP penalties are imposed on the coefficients associated with both scalar and functional covariates. The resulting penalized estimation problem is given by:

\begin{equation}\label{eqmcp}
\resizebox{0.9\textwidth}{!}{$
\displaystyle
(\hat{\boldsymbol{\beta}}(\tau), \hat{\boldsymbol{\alpha}}(\tau))
= \underset{\boldsymbol{\beta} (\tau), \boldsymbol{\alpha} (\tau)}{\arg\min}
\left\{
\frac{1}{2n}\Big(h_\gamma^\tau(Y_i -\eta_i) + \sum_{b=1}^{B} P_{\mathrm{MCP},\lambda,\phi}(\|\beta_b(\tau)\|_1)
+ \sum_{j=1}^{J} P_{\mathrm{MCP},\lambda_j,\phi}(\|\boldsymbol{\alpha}_j(\tau)\|_2)\Big)
\right\},
$}
\end{equation}
\begin{equation*}
P_{MCP,\lambda,\phi}(|\cdot|)=
\begin{cases}
\lambda|\cdot|-\frac{|\cdot|}{2\phi}\hspace{1.4 cm} &\text{if $|\cdot|\leq \lambda\phi$}.\\
.5\lambda^2\phi \hspace{2.7 cm} &\text{if $|\cdot|>\lambda\phi$}.
\end{cases}
\end{equation*}
Once $\hat{\boldsymbol{\alpha}}_j(\tau)$ is obtained, the estimated coefficient function can be reconstructed as $\hat{\Gamma}_{j}(s,\tau)
= \hat{\boldsymbol{\alpha}}_j(\tau)^{\top}\boldsymbol{\phi}_{j}(s)$, where $\boldsymbol{\phi}_{j}(s) = \bigl(\phi_{j1}(s), \ldots, \phi_{jQ_j}(s)\bigr)^{\top}$. The penalized estimation criterion in~(\ref{eqmcp}) for the proposed VSFLQR method can be efficiently optimized using a group descent algorithm \citep{yi2017semismooth}. The details of this algorithm are presented in Web Appendix A. We have used the \texttt{rqPen} package \citep{sherwood2025package} in R for performing the above optimization. We set $\lambda_j = \lambda \sqrt{G_j}$, where $G_j$ denotes the size of the group, to account for the different number of scores ($Q_j$) corresponding to the functional parameters.

\subsection{Choosing the tuning parameters}
Until now, we have assumed that the parameter \(\lambda\) as known. To determine the optimal tuning parameter, we use an  
extended Bayesian information criterion (EBIC) 
\citep{chen2008extended}. The proposed EBIC for a certain quantile is defined as $EBIC^{(\tau)}_{\lambda} = BIC^{(\tau)}_\lambda + 2 \log{p \choose \nu^{(\tau)}}$, where $\nu^{(\tau)}$ denotes the number of selected scalar and functional variables at quantile $\tau$, and $p$ denotes the total number of scalar and functional variables. Specifically, the BIC is computed as: $BIC^{(\tau)}_\lambda =  log \left[ \sum_{i=1}^n \rho_\tau \left(Y_i - \eta_i \right) \right] + log (n)*\nu^\tau$. Despite employing Huber approximations to improve numerical stability and computational efficiency, the selection of \(\lambda\) remains guided by the quantile loss function, ensuring robustness. The optimal value of $\lambda$ is then chosen using a grid search, producing the minimum EBIC. We set the MCP parameter $\phi=3$ and Huber loss function parameter $\gamma=0.2$, following the recommendations of the original authors \citep{zhang2010nearly,sherwood2022quantile}.

\section{Simulation Study}
\label{sec:sim_stud}
\subsection{Simulation Setup}
In this section, we assess the effectiveness of our variable selection method, VSFLQR, through numeric simulations. We generate data from a linear location-scale model \citep{kato2012estimation}. Specifically, we generate response values \( Y_i \) from the following model:
\[
Y_i = \boldsymbol{X_{i}}^{\top} \boldsymbol{\beta} + \sum_{j=1}^{20} \int_0^1 Z_{ij}(s) \Gamma_{j}(s) ds + \epsilon_i \left( X_{i1} \tilde{\beta}_1 + \int_0^1 Z_{i2}(s) \tilde{\Gamma}_{2}(s) ds \right),
\]
where \( \boldsymbol{\beta} = (\beta_0, \beta_1, \beta_2, \dots, \beta_{15})^{\top} \) represents the vector of scalar regression coefficients, and the set of scalar covariates is denoted as \( \boldsymbol{X_{i}} = (X_{i0}, X_{i1}, X_{i2}, \dots, X_{i15})^{\top} \). The error term $\epsilon_i$ is assumed to follow a Student's $t$-distribution with 5 degrees of freedom. Therefore, the quantile-dependent regression coefficients can be expressed as \( \boldsymbol{\beta}(\tau) = \boldsymbol{\beta} + Q_{\epsilon}(\tau)\tilde{\boldsymbol{\beta}} \), and the quantile-dependent functional coefficients are given by \( \Gamma_j(s,\tau) = \Gamma_j(s) + Q_{\epsilon}(\tau)\tilde{\Gamma}_j(s) \), where \( Q_{\epsilon}(\tau) \) denotes the quantile function associated with the error term \( \epsilon_i \). The regression coefficients are specified as follows: \( \beta_0 = 1 \), \( \beta_1 = 2 \), \( \beta_2 = 3 \), and \( \beta_3 = 4 \), while the remaining coefficients are set to zero, i.e., \( \beta_b = 0 \) for \( b = 4, 5, \dots, 15 \), indicating that the last 12 scalar covariates do not contribute to the model. The true underlying functional regression coefficients are specified as: \(\Gamma_{1}(s) = 3\cos(\pi s)\), \(\Gamma_{2}(s) = 4.5\sin(\pi s)\), \(\Gamma_{3}(s) = 3.5\cos(2\pi s) + 5.5\sin(-2\pi s)\), \(\Gamma_{4}(s) = 4\cos(2\pi s)\), and \(\Gamma_{5}(s) = 2.5\sin(2\pi s)\), with the remaining functional coefficients set to zero, i.e., \( \Gamma_j(s) = 0 \) for \( j = 6, 7, \dots, 20 \), implying that the last 15 functional covariates are irrelevant. We specify \( \tilde{\beta}_1 = 0.1 \) and \( \tilde{\Gamma}_{2}(s) = 0.1* \beta_{2}(s) \), which introduce heteroscedasticity in the response. The scalar covariates \( X_{ib} \) are independently drawn from different distributions: \( X_b = 1 \) if \( b = 0 \) (representing an intercept term); \( X_b \sim \text{Uniform}(0,1) \) for \( b = 1 \); \( X_b \sim \text{Uniform}(-1,1) \) for \( b \in \{2,3,4,\dots,15\} \). The functional covariates \( Z_{ij}(s) \) are generated using different basis function expansions: For \( j \in \{1,3,\dots,20\} \), we express each functional predictor as a linear combination of orthogonal basis polynomials: \(Z_{ij}(s) = \sum_{q=1}^{10} \omega_{ijq} \phi_{q}(s)\), where \( \phi_{q}(s) \) represent an orthonormal polynomial basis of degree \( 9 \), and the random coefficients \( \omega_{ijq} \) follow a normal distribution with mean zero and variance \( \sigma^2_q = 4q \). For \( j = 2 \), the functional predictor is generated using Gaussian basis functions: \(Z_{ij}(s) = \sum_{q=1}^{10} \omega_{ijq} \phi_{q}(s)\), where \( \phi_{q}(s) \) are Gaussian basis functions, and the corresponding coefficients \( \omega_{ijq} \) follow a \(\text{Uniform}(0,1) \) distribution. We generated $X_{i1}$ and $Z_{i2}(s)$ differently (ensuring they are positive) from the other covariates to guarantee identifiability of their corresponding regression coefficients (by ensuring the linear predictors are positive). We also considered dense and sparse sampling designs for the functional covariates.

\textbf{Scenario A: Dense Sampling}  

In this setting, functional predictors \( Z_{ij}(s) \) are recorded over a finely spaced, uniform grid of time points. Specifically, observations are taken at \( s = m/100 \) for \( m = 0,1,\dots,100 \), covering the entire interval \( [0,1] \).  For Scenario A, we examined three sample sizes ($n=200$, $n=400$, and $n=800$) at three quantile levels ($\tau=0.1$, $0.3$, and $0.5$) for assessing the estimation and selection performance. 

\textbf{Scenario B: Sparse Sampling}

Here, functional predictors \( Z_{ij}(s) \) are observed at a limited number of time points, selected randomly for each subject. The number of observations, \( m_{ij} \), is independently drawn from a uniform distribution, \( m_{ij} \stackrel{iid}{\sim} \text{Uniform}(20,31) \). These time points are then chosen from the same uniform grid used in Scenario A, i.e., \( s = m/100 \) for \( m = 0,1,\dots,100 \). Under Scenario B, the sample size configurations are varied by quantile level: for $\tau=0.5$, we considered $n=200$, $400$, and $800$; for $\tau=0.3$, we used $n=400$, $800$, and $1600$; and for $\tau=0.1$, the sample sizes were $n=800$, $1600$, and $3200$. For each of these configurations, we generated 200 Monte Carlo datasets ($n_d = 200$) to assess the performance of our proposed approach. An independent test split (sample size $25\%$ of training size) was used in each replication for assessing the prediction performance in both scenarios.

\subsection{Simulation results}
Our primary objective is to identify important scalar and functional predictors while accurately estimating the scalar coefficients $\boldsymbol{\beta}(\tau)$ and the functional coefficient curves $\Gamma_j(s,\tau)$ across different quantile levels. We evaluate the performance of the proposed VSFLQR by comparing it with two competing methods: i) quantile regression with group Lasso (rqgLasso) and ii) mean regression with group Lasso (grpregLasso) \citep{breheny2015group}. However, grpregLasso is designed for mean regression and is therefore only applicable at $\tau = 0.5$. The tuning parameters for all methods are selected automatically using the proposed EBIC criterion. 
Table~\ref{tqs} summarizes the variable selection performance under Scenario~A. Specifically, it reports the true positive rate (TPR) and false positive rate (FPR) for scalar variables, functional variables, and all variables combined, across three sample sizes and three quantile levels. The table also presents the average model size. Overall, VSFLQR demonstrates consistently strong selection performance, successfully identifying relevant predictors while excluding irrelevant ones. Both VSFLQR and grpregLASSO achieve high TPRs, indicating accurate recovery of the true signals. However, only VSFLQR maintains a low FPR for functional covariates and scalar covariates, highlighting its robustness in eliminating irrelevant features. These results underscore the effectiveness of VSFLQR for joint scalar and functional variable selection, particularly as the sample size increases.

\begin{sidewaystable}[h!]
\centering
{
\caption{Comparison of selection performance across different sample sizes and quantile levels (dense design).}
\label{tqs}
\resizebox{\textwidth}{!}{%
\begin{tabular}{cccccccccc}
\hline
Sample size & Method      & Quantile & \begin{tabular}[c]{@{}c@{}}TPR of\\ scalar variables\end{tabular} & \begin{tabular}[c]{@{}c@{}}FPR of\\ scalar variables\end{tabular} & \begin{tabular}[c]{@{}c@{}}TPR of\\ functional variables\end{tabular} & \begin{tabular}[c]{@{}c@{}}FPR of\\ functional variables\end{tabular} & \begin{tabular}[c]{@{}c@{}}TPR of\\ all variables\end{tabular} & \begin{tabular}[c]{@{}c@{}}FPR of\\ all variables\end{tabular} & Average model size \\ \hline
200         & grpregLasso & 0.5      & 1.000                                                             & 0.148                                                             & 1.000                                                                 & 0.042                                                                 & 1.000                                                          & 0.089                                                          & 10.405             \\
            &             &          &                                                                   &                                                                   &                                                                       &                                                                       &                                                                &                                                                &                    \\
            & rqgLasso    & 0.5      & 0.722                                                             & 0.006                                                             & 0.154                                                                 & 0.000                                                                 & 0.367                                                          & 0.003                                                          & 3.010              \\
            &             &          &                                                                   &                                                                   &                                                                       &                                                                       &                                                                &                                                                &                    \\
            &             & 0.3      & 0.685                                                             & 0.002                                                             & 0.015                                                                 & 0.000                                                                 & 0.266                                                          & 0.001                                                          & 2.150              \\
            &             &          &                                                                   &                                                                   &                                                                       &                                                                       &                                                                &                                                                &                    \\
            &             & 0.1      & 0.698                                                             & 0.018                                                             & 0.111                                                                 & 0.000                                                                 & 0.331                                                          & 0.008                                                          & 2.860              \\
            &             &          &                                                                   &                                                                   &                                                                       &                                                                       &                                                                &                                                                &                    \\
            & VSFLQR      & 0.5      & 1.000                                                             & 0.005                                                             & 1.000                                                                 & 0.000                                                                 & 1.000                                                          & 0.002                                                          & 8.060              \\
            &             &          &                                                                   &                                                                   &                                                                       &                                                                       &                                                                &                                                                &                    \\
            &             & 0.3      & 1.000                                                             & 0.003                                                             & 1.000                                                                 & 0.000                                                                 & 1.000                                                          & 0.001                                                          & 8.035              \\
            &             &          &                                                                   &                                                                   &                                                                       &                                                                       &                                                                &                                                                &                    \\
            &             & 0.1      & 1.000                                                             & 0.016                                                             & 1.000                                                                 & 0.002                                                                 & 1.000                                                          & 0.008                                                          & 8.220              \\
            &             &          &                                                                   &                                                                   &                                                                       &                                                                       &                                                                &                                                                &                    \\
400         & grpregLasso & 0.5      & 1.000                                                             & 0.109                                                             & 1.000                                                                 & 0.015                                                                 & 1.000                                                          & 0.057                                                          & 9.535              \\
            &             &          &                                                                   &                                                                   &                                                                       &                                                                       &                                                                &                                                                &                    \\
            & rqgLasso    & 0.5      & 0.693                                                             & 0.000                                                             & 0.152                                                                 & 0.000                                                                 & 0.355                                                          & 0.000                                                          & 2.840              \\
            &             &          &                                                                   &                                                                   &                                                                       &                                                                       &                                                                &                                                                &                    \\
            &             & 0.3      & 0.670                                                             & 0.000                                                             & 0.006                                                                 & 0.000                                                                 & 0.255                                                          & 0.000                                                          & 2.040              \\
            &             &          &                                                                   &                                                                   &                                                                       &                                                                       &                                                                &                                                                &                    \\
            &             & 0.1      & 0.687                                                             & 0.003                                                             & 0.062                                                                 & 0.000                                                                 & 0.296                                                          & 0.001                                                          & 2.400              \\
            &             &          &                                                                   &                                                                   &                                                                       &                                                                       &                                                                &                                                                &                    \\
            & VSFLQR      & 0.5      & 1.000                                                             & 0.000                                                             & 1.000                                                                 & 0.000                                                                 & 1.000                                                          & 0.000                                                          & 8.000              \\
            &             &          &                                                                   &                                                                   &                                                                       &                                                                       &                                                                &                                                                &                    \\
            &             & 0.3      & 1.000                                                             & 0.000                                                             & 1.000                                                                 & 0.000                                                                 & 1.000                                                          & 0.000                                                          & 8.000              \\
            &             &          &                                                                   &                                                                   &                                                                       &                                                                       &                                                                &                                                                &                    \\
            &             & 0.1      & 1.000                                                             & 0.002                                                             & 1.000                                                                 & 0.000                                                                 & 1.000                                                          & 0.001                                                          & 8.020              \\
            &             &          &                                                                   &                                                                   &                                                                       &                                                                       &                                                                &                                                                &                    \\
800         & grpregLasso & 0.5      & 1.000                                                             & 0.081                                                             & 1.000                                                                 & 0.006                                                                 & 1.000                                                          & 0.039                                                          & 9.060              \\
            &             &          &                                                                   &                                                                   &                                                                       &                                                                       &                                                                &                                                                &                    \\
            & rqgLasso    & 0.5      & 0.678                                                             & 0.000                                                             & 0.154                                                                 & 0.000                                                                 & 0.351                                                          & 0.000                                                          & 2.805              \\
            &             &          &                                                                   &                                                                   &                                                                       &                                                                       &                                                                &                                                                &                    \\
            &             & 0.3      & 0.670                                                             & 0.000                                                             & 0.000                                                                 & 0.000                                                                 & 0.251                                                          & 0.000                                                          & 2.010              \\
            &             &          &                                                                   &                                                                   &                                                                       &                                                                       &                                                                &                                                                &                    \\
            &             & 0.1      & 0.677                                                             & 0.000                                                             & 0.038                                                                 & 0.000                                                                 & 0.278                                                          & 0.000                                                          & 2.220              \\
            &             &          &                                                                   &                                                                   &                                                                       &                                                                       &                                                                &                                                                &                    \\
            & VSFLQR      & 0.5      & 1.000                                                             & 0.000                                                             & 1.000                                                                 & 0.000                                                                 & 1.000                                                          & 0.000                                                          & 8.000              \\
            &             &          &                                                                   &                                                                   &                                                                       &                                                                       &                                                                &                                                                &                    \\
            &             & 0.3      & 1.000                                                             & 0.000                                                             & 1.000                                                                 & 0.000                                                                 & 1.000                                                          & 0.000                                                          & 8.000              \\
            &             &          &                                                                   &                                                                   &                                                                       &                                                                       &                                                                &                                                                &                    \\
            &             & 0.1      & 1.000                                                             & 0.000                                                             & 1.000                                                                 & 0.000                                                                 & 1.000                                                          & 0.000                                                          & 8.000              \\ \hline
\end{tabular}%
}
}
\end{sidewaystable}

Beyond selection performance, VSFLQR also achieves strong estimation accuracy across different quantile levels. It consistently yields minimal bias and mean squared error (MSE) for the scalar parameters and low mean integrated squared error (MISE) for the functional parameters reported in Table \ref{tqe}. The MISE of the functional estimate \(\hat{\Gamma}_j(s,\tau)\) is defined as $
MISE_{(j,\tau)} = \frac{1}{n_d}\sum_{d=1}^{n_d} \left( \int_0^1 \left( \hat{\Gamma}_{jd}(s,\tau) - \Gamma_j(s,\tau) \right)^2 \, ds \right)$,
where \(\hat{\Gamma}_{jd}(s,\tau)\) is the estimate of \(\Gamma_j(s,\tau)\) for the \(d\)-th generated dataset, and \(n_d\) is the number of datasets. Compared to rqgLasso, VSFLQR reduces the MSE of scalar parameter estimates by an order of magnitude or more across all quantile levels, and achieves MISE reductions of a similar or greater scale for the functional coefficients. Relative to grpregLasso at $\tau = 0.5$, VSFLQR yields scalar MSE values roughly 5 times smaller and functional MISE values approximately 2 to 3 times smaller on average. As the sample size increases, the magnitude of VSFLQR's MSE and MISE steadily decreases, demonstrating its efficiency in the estimation of the scalar and functional coefficients.

\begin{sidewaystable}[h!]
\centering
{
\caption{Comparison of Bias, MSE, and MISE across different sample sizes and quantile levels (dense design).}
\label{tqe}
\resizebox{\textwidth}{!}{%
\begin{tabular}{cccccccccccccccccccccc}
\hline
Sample size & Method      & Quantile &  & \multicolumn{2}{c}{$\hat{\beta_1}$} &  & \multicolumn{2}{c}{$\hat{\beta_2}$} &  & \multicolumn{2}{c}{$\hat{\beta_3}$} &  & $\hat{\Gamma}_1(\cdot, \tau)$ &  & $\hat{\Gamma}_2(\cdot, \tau)$ &  & $\hat{\Gamma}_3(\cdot, \tau)$ &  & $\hat{\Gamma}_4(\cdot, \tau)$ &  & $\hat{\Gamma}_5(\cdot, \tau)$ \\ \cline{5-6} \cline{8-9} \cline{11-12} \cline{14-14} \cline{16-16} \cline{18-18} \cline{20-20} \cline{22-22} 
            &             &          &  & Bias              & MSE             &  & Bias              & MSE             &  & Bias              & MSE             &  & MISE                          &  & MISE                          &  & MISE                          &  & MISE                          &  & MISE                          \\ \cline{1-3} \cline{5-6} \cline{8-9} \cline{11-12} \cline{14-14} \cline{16-16} \cline{18-18} \cline{20-20} \cline{22-22} 
200         & grpregLasso & 0.5      &  & -0.205            & 0.065           &  & -0.111            & 0.019           &  & -0.111            & 0.019           &  & 0.157                         &  & 1.016                         &  & 0.217                         &  & 0.184                         &  & 0.172                         \\
            &             &          &  &                   &                 &  &                   &                 &  &                   &                 &  &                               &  &                               &  &                               &  &                               &  &                               \\
            & rqgLasso    & 0.5      &  & -1.834            & 3.533           &  & -1.283            & 1.958           &  & -1.312            & 2.137           &  & 4.545                         &  & 10.025                        &  & 12.831                        &  & 8.079                         &  & 3.094                         \\
            &             & 0.3      &  & -1.864            & 3.576           &  & -1.815            & 3.678           &  & -1.842            & 3.767           &  & 4.545                         &  & 8.934                         &  & 20.412                        &  & 8.079                         &  & 3.094                         \\
            &             & 0.1      &  & -1.705            & 3.131           &  & -1.738            & 3.533           &  & -1.824            & 3.925           &  & 4.545                         &  & 7.268                         &  & 16.198                        &  & 8.057                         &  & 3.094                         \\
            &             &          &  &                   &                 &  &                   &                 &  &                   &                 &  &                               &  &                               &  &                               &  &                               &  &                               \\
            & VSFLQR      & 0.5      &  & 0.004             & 0.012           &  & 0.001             & 0.004           &  & -0.003            & 0.004           &  & 0.072                         &  & 0.445                         &  & 0.090                         &  & 0.082                         &  & 0.054                         \\
            &             & 0.3      &  & 0.020             & 0.015           &  & -0.006            & 0.004           &  & 0.001             & 0.003           &  & 0.072                         &  & 0.434                         &  & 0.097                         &  & 0.085                         &  & 0.062                         \\
            &             & 0.1      &  & 0.067             & 0.034           &  & -0.008            & 0.005           &  & 0.001             & 0.005           &  & 0.098                         &  & 0.650                         &  & 0.114                         &  & 0.102                         &  & 0.108                         \\
            &             &          &  &                   &                 &  &                   &                 &  &                   &                 &  &                               &  &                               &  &                               &  &                               &  &                               \\
400         & grpregLasso & 0.5      &  & -0.140            & 0.030           &  & -0.075            & 0.008           &  & -0.076            & 0.008           &  & 0.068                         &  & 0.403                         &  & 0.107                         &  & 0.081                         &  & 0.076                         \\
            &             &          &  &                   &                 &  &                   &                 &  &                   &                 &  &                               &  &                               &  &                               &  &                               &  &                               \\
            & rqgLasso    & 0.5      &  & -1.912            & 3.760           &  & -1.250            & 1.793           &  & -1.340            & 2.079           &  & 4.545                         &  & 10.025                        &  & 12.672                        &  & 8.079                         &  & 3.094                         \\
            &             & 0.3      &  & -1.939            & 3.763           &  & -1.638            & 2.892           &  & -1.684            & 3.053           &  & 4.545                         &  & 8.934                         &  & 20.725                        &  & 8.079                         &  & 3.094                         \\
            &             & 0.1      &  & -1.807            & 3.322           &  & -1.720            & 3.293           &  & -1.891            & 4.020           &  & 4.545                         &  & 7.284                         &  & 17.588                        &  & 8.079                         &  & 3.094                         \\
            &             &          &  &                   &                 &  &                   &                 &  &                   &                 &  &                               &  &                               &  &                               &  &                               &  &                               \\
            & VSFLQR      & 0.5      &  & -0.005            & 0.007           &  & -0.005            & 0.002           &  & -0.002            & 0.001           &  & 0.041                         &  & 0.168                         &  & 0.051                         &  & 0.053                         &  & 0.023                         \\
            &             & 0.3      &  & 0.020             & 0.008           &  & 0.000             & 0.002           &  & -0.007            & 0.002           &  & 0.044                         &  & 0.165                         &  & 0.055                         &  & 0.056                         &  & 0.023                         \\
            &             & 0.1      &  & 0.081             & 0.016           &  & -0.001            & 0.002           &  & -0.009            & 0.002           &  & 0.051                         &  & 0.281                         &  & 0.062                         &  & 0.063                         &  & 0.033                         \\
            &             &          &  &                   &                 &  &                   &                 &  &                   &                 &  &                               &  &                               &  &                               &  &                               &  &                               \\
800         & grpregLasso & 0.5      &  & -0.102            & 0.015           &  & -0.049            & 0.004           &  & -0.049            & 0.003           &  & 0.039                         &  & 0.200                         &  & 0.059                         &  & 0.050                         &  & 0.038                         \\
            &             &          &  &                   &                 &  &                   &                 &  &                   &                 &  &                               &  &                               &  &                               &  &                               &  &                               \\
            & rqgLasso    & 0.5      &  & -1.966            & 3.902           &  & -1.281            & 1.851           &  & -1.329            & 2.011           &  & 4.545                         &  & 10.025                        &  & 12.700                        &  & 8.079                         &  & 3.094                         \\
            &             & 0.3      &  & -1.937            & 3.759           &  & -1.615            & 2.701           &  & -1.627            & 2.733           &  & 4.545                         &  & 8.934                         &  & 21.161                        &  & 8.079                         &  & 3.094                         \\
            &             & 0.1      &  & -1.835            & 3.383           &  & -1.768            & 3.348           &  & -1.963            & 4.174           &  & 4.545                         &  & 7.284                         &  & 18.786                        &  & 8.079                         &  & 3.094                         \\
            &             &          &  &                   &                 &  &                   &                 &  &                   &                 &  &                               &  &                               &  &                               &  &                               &  &                               \\
            & VSFLQR      & 0.5      &  & -0.007            & 0.003           &  & -0.003            & 0.001           &  & -0.002            & 0.001           &  & 0.029                         &  & 0.086                         &  & 0.036                         &  & 0.041                         &  & 0.011                         \\
            &             & 0.3      &  & 0.027             & 0.003           &  & 0.003             & 0.001           &  & -0.001            & 0.001           &  & 0.029                         &  & 0.092                         &  & 0.039                         &  & 0.040                         &  & 0.011                         \\
            &             & 0.1      &  & 0.087             & 0.011           &  & 0.004             & 0.001           &  & -0.001            & 0.001           &  & 0.032                         &  & 0.165                         &  & 0.042                         &  & 0.043                         &  & 0.014                         \\ \hline
\end{tabular}%
}
}
\end{sidewaystable}
We present the Monte Carlo means of the estimated functional coefficients $\hat{\Gamma}_j(s,\tau)$ ($j = 1,2,3,4,5$) obtained from the proposed method, overlaid on the true regression curves, in Figure~\ref{fig05qs400} for $n = 400$ and $\tau=0.5$. Pointwise 95\% Monte Carlo confidence intervals (CIs) are also included. These CIs are constructed using a percentile-based approach, taking the pointwise 2.5th and 97.5th percentiles of the estimated coefficient functions across all replications. The VSFLQR estimates closely follow the true curves, effectively capturing the underlying functional relationships. The findings for $\tau=0.3$ and $\tau=0.1$ show comparable trends and are presented in Web Figures 1, 2. 

Table \ref{tqp} summarizes the average mean squared prediction error (MSPE) and average mean absolute prediction error (MAPE) for $\tau=0.5$ for the three methods on the test dataset. Across all settings, VSFLQR attains the lowest MSPE and MAPE values. These results indicate that, in addition to strong selection and estimation performance, VSFLQR also provides accurate and reliable predictions. Additional simulation results for the sparse functional data are provided in the supporting materials (Web Tables 1-3 and Web Figures 3-5). The main conclusions remain consistent across different quantile levels, with some degradation under the sparse sampling, which is expected.

\newcommand{\solidline}{\raisebox{2pt}{\tikz{\draw[-,black,solid,line width = 0.3pt](0,0) -- (5mm,0);}}}
\begin{figure}[H]
    \centering
    \includegraphics[scale=0.5]{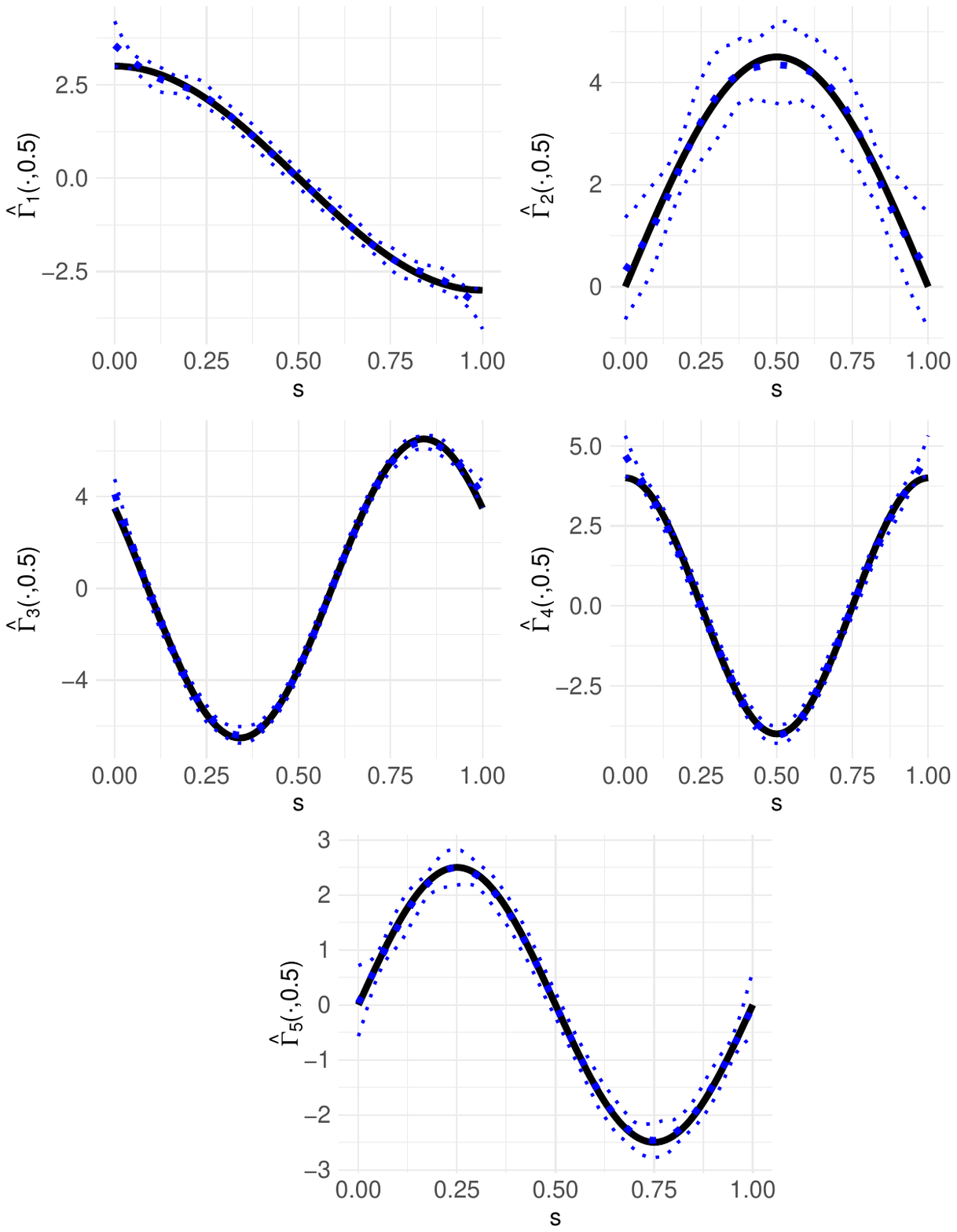}
    \caption{MC estimates and pointwise confidence intervals of the coefficient functions ($\tau=0.5$, $n=400$, dense design); (\textcolor{blue}{\textbf{$\cdots$}}, VSFCOX; \protect\solidline, true curve).}
    \label{fig05qs400}
\end{figure}

VSFLQR consistently produces higher estimation, selection, and prediction accuracy compared to the other two methods, highlighting its usefulness. Overall, these findings indicate that VSFLQR not only performs well in variable selection but also yields highly accurate estimates of both scalar and functional coefficients across the different quantile levels thus facilitating identification and accurate estimation of heterogeneous covariate effects under both dense and sparse designs.

\begin{table}[H]
\centering
\caption{Comparison of prediction performance across different sample sizes (dense design). Displayed are the average mean squared prediction error (MSPE) and average mean absolute prediction error (MAPE) for $\tau=0.5$ on the test dataset.}
\begin{tabular}{cccc}
\hline
Sample size & Method      & MSPE   & MAPE  \\ \hline
200         & grpregLasso & 0.456  & 0.526 \\
            &             &        &       \\
            & rqgLasso    & 11.568 & 2.703 \\
            &             &        &       \\
            & VSFLQR      & 0.279  & 0.396 \\
            &             &        &       \\
400         & grpregLasso & 0.326  & 0.431 \\
            &             &        &       \\
            & rqgLasso    & 11.390 & 2.673 \\
            &             &        &       \\
            & VSFLQR      & 0.255  & 0.370 \\
            &             &        &       \\
800         & grpregLasso & 0.279  & 0.393 \\
            &             &        &       \\
            & rqgLasso    & 11.471 & 2.690 \\
            &             &        &       \\
            & VSFLQR      & 0.245  & 0.361 \\ \hline
\end{tabular}
\label{tqp}
\end{table}

\section{Real data application: Modelling Cognitive Function in NHANES 2011-2014}
\label{realdat}
We apply the VSFLQR method to accelerometer data from the NHANES 2011--2014 waves to identify key temporally varying distributional patterns of physical activity, along with demographic predictors, that are associated with cognitive function, capturing heterogeneous associations across different quantiles. Although numerous studies have examined the association between PA and cognitive functioning, most have relied on scalar summary measures such as total activity count (TAC), moderate-to-vigorous PA (MVPA), or mean \citep{campbell2023estimating,quinlan2023physical,yau2025physical}. While these scalar metrics facilitate interpretation, they fail to capture the full range of temporal variation in PA intensity \citep{ghosal2023distributional}. Because physical activity patterns can vary substantially over the course of a day, preserving temporal information is essential for characterizing the circadian rhythm of PA \citep{xiao2015quantifying} and for understanding its implications for cognitive function. Recent studies \citep{ghosal2022scalar} have further shown that distributional features beyond mean PA, including higher-order moments such as variability and skewness, yield valuable and complementary information. In this regard, time-of-day-dependent daily L-moments \citep{ghosal2022scalar, ghosal2023distributional, cho2024exploring} provide an effective framework for capturing the daily distributional patterns of physical activity.

The National Health and Nutrition Examination Survey (NHANES) provides nationally representative health and nutrition data for the civilian, noninstitutionalized U.S.\ population. During the 2011--2014 survey waves, physical activity data were collected using the wrist-worn ActiGraph GT3X+ accelerometer (ActiGraph, Pensacola, FL). Participants were instructed to wear the monitor continuously for seven full days and to remove it on the morning of the ninth day. Our analysis uses the 2011--2014 minute-level accelerometer data, made publicly available in 2021. These data record physical activity in Monitor-Independent Movement Summary (MIMS) units, an open-source and device-independent metric designed to provide a standardized measure of PA \citep{john2019open}.
As mentioned in the introduction, we consider the triaxial minute-level MIMS summary measure as the primary exposure, and cognitive scores from the animal fluency test (CFDAST) as our outcome. The objective of our analysis is to understand how the conditional distribution of the cognitive function depends on daily physical activity patterns beyond the daily mean activity and identify the influential daily PA patterns together with the key demographic and lifestyle variables. In NHANES 2011-2014, the cognitive measures were obtained through a set of standardized assessments administered to participants aged 60 years and above. Our final analytic sample consists of 2,393 adults aged 60 years and older who had available cognitive scores and at least four days of valid accelerometer data with a minimum of ten hours of wear time per day. Descriptive statistics for the full sample are provided in Web Table 4. Web Figure 6 displays the distribution of CFDAST scores. The scalar covariates considered in the analysis were age ($X_{1}$), body mass index (BMI; $X_{2}$), gender ($X_{3}$), the ratio of family income to poverty ($X_{4}$), and education level, categorized as high school or less (reference group), and beyond high school ($X_{5}$).

To capture how the distribution of PA evolves over the course of a day, we use daily time-of-day-specific L-moments as functional exposure variables \citep{ghosal2022scalar,yue2026variable}. L-moments are rank-based counterparts of traditional moments \citep{hosking1990moments, ghosal2023distributional}. They summarize key aspects of a distribution such as location, spread, asymmetry, and tail behavior while being less affected by extreme observations than classical moment-based measures \citep{vogel1993moment}. This feature is particularly useful for PA count data, which are typically skewed and contain a large proportion of zeros. Empirically, L-moments are constructed as linear combinations of order statistics, and thus belong to the class of L-statistics. The first four L-moments correspond to measures analogous to the mean, variability, skewness, and kurtosis. These are denoted by $L_1$ (equal to the mean), $L_2$ (L-scale), $L_3$ (L-skewness), and $L_4$ (L-kurtosis), respectively. In general, the $r$th population L-moment of a random variable $X$ is given by
\[
L_r = \frac{1}{r} \sum_{k=0}^{r-1} (-1)^k \binom{r-1}{k} \, \mathbb{E}\!\left(X_{r-k:r}\right), \qquad r = 1,2,\ldots,
\]
where $X_{1:n} \le X_{2:n} \le \cdots \le X_{n:n}$ denote the order statistics from a random sample of size $n$ drawn from the distribution of $X$.

For subject $i=1,\ldots,n$, let $X_{ik}(s)$ denote the log-transformed MIMS recorded at minute-level resolution on day $k$ at time-of-day $s$, where $k=1,\ldots,n_i$. Measurements are taken over the grid \(S=\left\{ \frac{m}{60} : m=0,1,\ldots,1439 \right\}\), which represents the 24-hour day in hours. To describe how the distribution of PA changes throughout the day, we construct the first four diurnal L-moment curves, denoted by $L_{ij}(s)$ for $j=1,2,3,4$ \citep{ghosal2022scalar}. At each time point $s$, the quantity $L_{ij}(s)$ is computed from the collection $\{X_{ik}(u)\}_{k=1}^{n_i}$ within a local window $u \in (s-\zeta, s+\zeta)$, where $\zeta=5/60$ hour (i.e., a 5-minute window). These L-moment trajectories preserve time-varying daily distributional patterns of PA, offering a richer representation than conventional summary indices \citep{varma2017re, cho2024exploring, yue2026variable}. 

Figure~\ref{fig1} (Panel b) shows the smoothed population averages together with curves from five randomly selected individuals. For notational simplicity, we set $Z_{ij}(s)=L_{ij}(s)$ for $j=1,2,3,4$. To assess whether the association between these distributional features and the outcome varies by age or gender, we further include interaction functions: $Z_{ij}(s)$ for $j=5,\ldots,8$ correspond to age-by–L-moment terms, and $j=9,\ldots,12$ correspond to gender-by–L-moment terms. In total, the model includes five scalar covariates and twelve functional predictors. The functional linear quantile regression is written as
\begin{equation*}
 Q_{Y_i} (\tau | X_{i1},\ldots, X_{i5}, Z_{i1}(\cdot), \dots, Z_{i12}(\cdot)) = \sum_{b=1}^{5} X_{ib}\beta_b(\tau) + \sum_{j=1}^{12} \int_0^{24} Z_{ij}(s) \Gamma_j(s, \tau) ds
\end{equation*}

\begin{figure}[h!]
    \centering
    \includegraphics[scale=0.4]{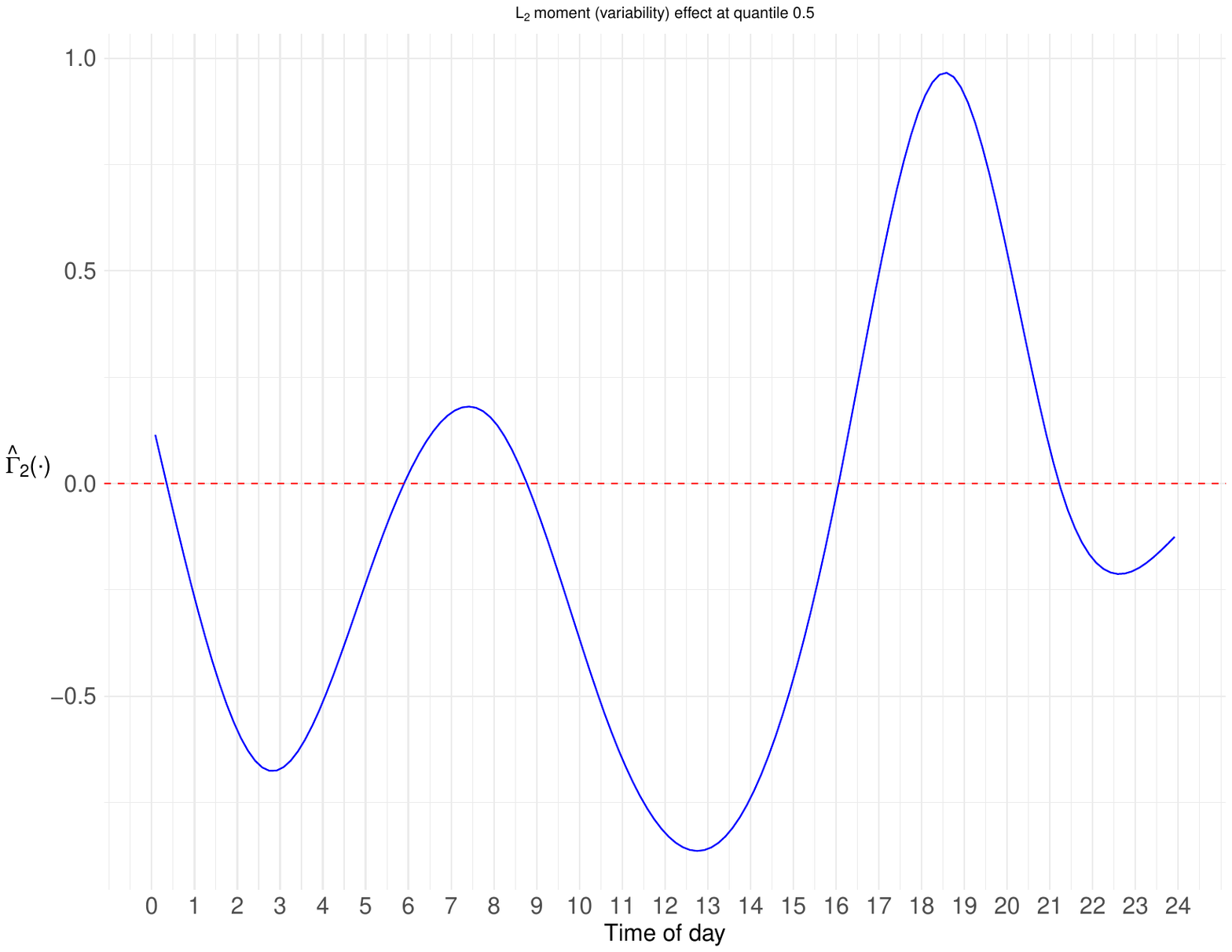}
    \caption{Estimated functional effects of the second ($L_2$) moment at $\tau=0.5$ over time.}
    \label{figLest}
\end{figure}

We fitted the functional linear quantile regression at $\tau \in \{0.1, 0.5, 0.9\}$ using the proposed VSFLQR method to simultaneously carry out variable selection and coefficient estimation for both scalar and functional predictors. Among the functional candidates, only the second L-moment curve was retained at $\tau=0.5$; the remaining L-moments and all interaction terms were excluded at all quantile levels. Figure~\ref{figLest} presents the estimated coefficient function $\hat{\Gamma}_2(\cdot)$ corresponding to the second L-moment, $L_2(\cdot)$, which captures the variability of PA throughout the day. The pattern of $\hat{\Gamma}_2(s)$ indicates that a higher contrast in activity levels during the early morning hours (approximately 6:00~a.m.\ to 9:00~a.m.) and late afternoon and evening (about 4:00~p.m.\ to 9:00~p.m.), compared to midday and night, is associated with a higher cognitive performance. These findings point to the importance of daily activity reserve in relation to cognitive health. In particular, maintaining greater variability in activity during specific hours may help protect against cognitive decline in older adults \citep{cho2024exploring, donahue2025activity}, potentially reflecting a larger functional reserve of PA. For persons already at the high or low end of the cognitive spectrum ($\tau=0.1,0.9$), the PA patterns were not found to be a key contributing factor for cognitive performance after adjusting for age, sex, and education.

The scalar variables retained by the proposed method at different quantile levels and their corresponding effects are summarized in Table~\ref{ts}. The selection procedure identified age, ratio of family income to poverty, and education level (beyond high school) as important predictors across the three different quantiles. Advancing age was associated with a decline in cognitive function \citep{yang2023cognitive,matsui2026decline}. In contrast, a higher ratio of family income to poverty \citep{mani2013poverty,krueger2025lifetime}, as well as a higher level of education \citep{clouston2020education,seyedsalehi2023educational}, were associated with better cognitive function. The estimated effects of the key scalar covariates increase in magnitude as the quantile level rises from 0.1 to 0.9, highlighting the heterogeneous effect of these covariates.

\begin{table}[H]
\centering
\caption{Selected scalar variables and corresponding effects across different quantile levels in the NHANES application.}
\resizebox{\textwidth}{!}{\begin{tabular}{ccccc}
\hline
           & Quantile & Age    & Ratio of family income to poverty & Education level (Beyond high school) \\ \hline
Estimation & 0.1      & -0.099 & 0.365                             & 2.092              \\
           & 0.5      & -0.139 & 0.485                             & 2.371              \\
           & 0.9      & -0.211 & 0.548                             & 3.306              \\ \hline
\end{tabular}}
\label{ts}
\end{table}

\begin{figure}[H]
    \centering
    \includegraphics[width=0.8\textwidth,height=0.6\textwidth]{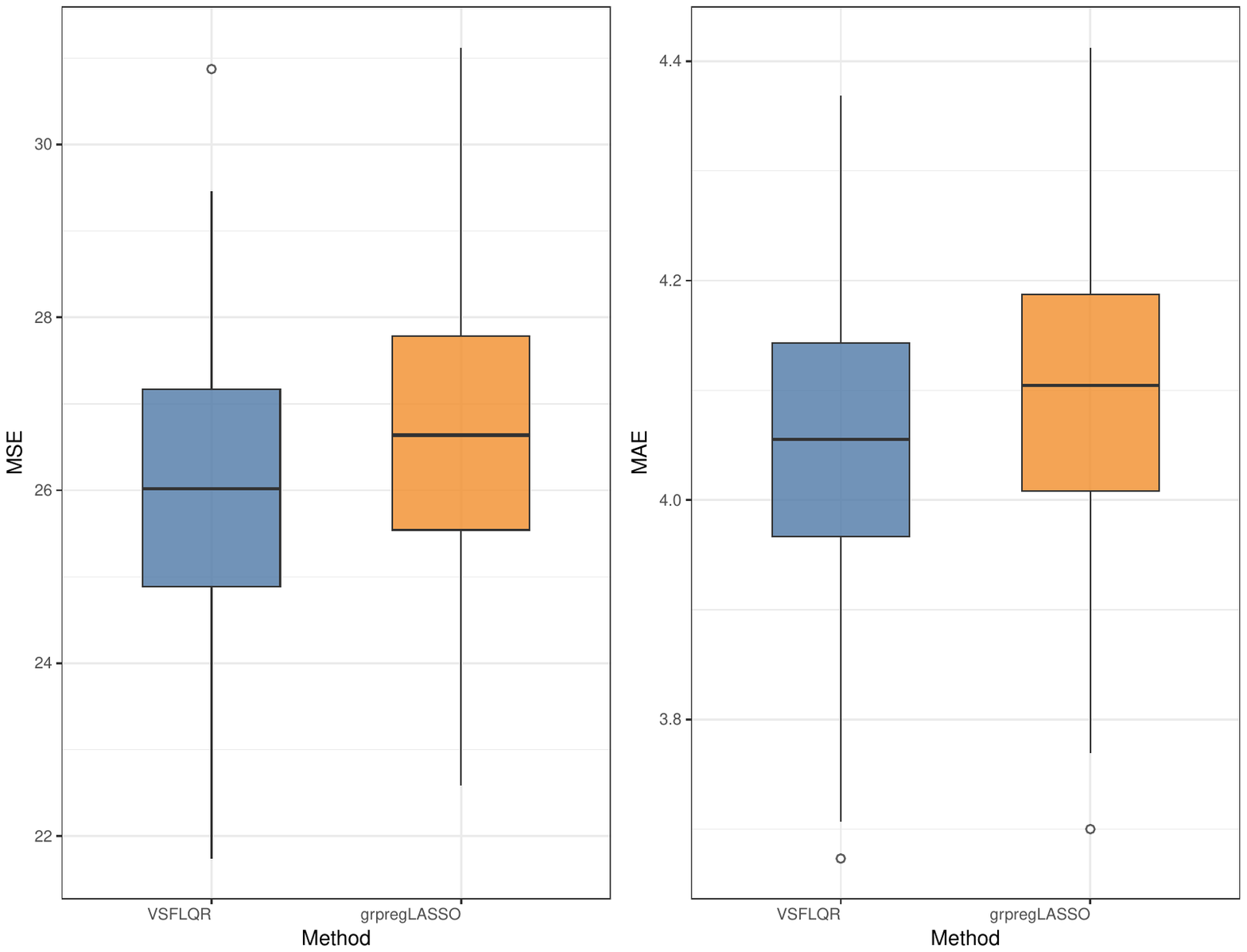}
    \caption{Comparison of prediction performance between VSFLQR and grpregLASSO in terms of MSE (left) and MAE (right).}
    \label{figboxplot}
\end{figure}

To assess the predictive performance at $\tau=0.5$, we computed the out-of-sample MSPE and MAPE for the proposed VSFLQR method and compared the results with those from grpregLASSO (performing mean regression). In each replication, 80\% of the data were randomly selected for model fitting, and prediction errors were evaluated on the remaining 20\%. This train-test split was repeated 100 times. As shown in Figure~\ref{figboxplot}, VSFLQR consistently produces lower prediction errors under both metrics. The boxplots reveal smaller median MSE and MAE values for VSFLQR relative to grpregLASSO. Overall, these results indicate that VSFLQR delivers more accurate and stable predictions in repeated sampling experiments.

As a further check on the variable selection performance, we augmented the dataset with 10 functional pseudo-predictors \citep{wu2007controlling, ghosal2020variable}. These pseudo-covariates were constructed to act as pure noise, allowing us to evaluate the specificity of the proposed method and its tendency to select irrelevant variables in a high-dimensional setting. For each subject, we generated $Z_{ij}(\cdot) \stackrel{iid}{\sim} Z_j(\cdot)$, where $j=13,14,\ldots,22$, and \(Z_j(h) = a_j \sqrt{10}\sin\!\left(\frac{\pi j h}{24}\right) 
+ b_j \sqrt{10}\cos\!\left(\frac{\pi j h}{24}\right)\), with $a_j \sim \mathcal{N}(0, 10^2)$ and $b_j \sim \mathcal{N}(0, 10^2)$. This yields a total of 27 covariates: the original 17 predictors and 10 simulated functional noise variables. We then applied the VSFLQR method using the observed outcomes $(Y_i)$ together with all scalar and functional covariates, namely $X_{i1}, X_{i2}, X_{i3}, \ldots, Z_{i1}(s), Z_{i2}(s), \ldots, Z_{i22}(s)$. The entire procedure was replicated 100 times, and we recorded how often each variable was selected across different quantile levels. The resulting selection frequencies are summarized in Web Table 5. Age, ratio of family income to poverty, and education level (beyond high school) were selected 100\% of the time at different quantile levels, underscoring their consistent association with cognitive function. At the 0.5 quantile, the $L_2(\cdot)$ moment was selected 88\% of the time. The other original covariates had a selection percentage of 5.3\%, while the pseudo-functional variables had a selection percentage of 1.7\%. These findings provide additional support for the ability of VSFLQR to reliably recover the key predictors while effectively excluding noise variables.

\section{Discussion}
\label{disc}
In this paper, we developed a variable selection approach for the functional linear quantile regression that accommodates both scalar and multiple functional predictors. The proposed method, VSFLQR, combines FPCA with a group MCP penalty, allowing the model to impose smoothness on coefficient functions while simultaneously encouraging sparsity in variable selection. Simulation studies demonstrate that VSFLQR achieves reliable identification of relevant predictors and accurate coefficient estimation, with performance improving as the sample size increases under a range of scenarios. When applied to the NHANES 2011--2014 data, the method yields meaningful public health insights regarding how daily activity patterns relate to cognitive function in older U.S.\ adults across different quantile levels. In particular, the second L-moment of physical activity, $L_2(\cdot)$, together with age, ratio of family income to poverty, and education level, emerged as the primary contributors to cognitive function at the median. These findings suggest that the day-to-day variability in physical activity plays an important role in cognitive function. Accounting for both timing \citep{feng2023associations} and intensity patterns \citep{cho2024exploring,donahue2025activity} may therefore help inform more targeted physical activity recommendations for older adults.

There are several directions in which the proposed VSFLQR framework could be further developed. One natural extension is to functional additive quantile regression \citep{horowitz2005nonparametric, waldmann2013bayesian,zhang2021functional}, which would permit nonlinear additive effects of both scalar and functional predictors. The current approach estimates models independently at each quantile level. Joint modeling across multiple quantiles, for example using quantile process regression \citep{koenker2005quantile} or composite quantile methods \citep{zou2008composite}, may improve efficiency and ensure smoothness across $\tau$. An important extension is to accommodate longitudinal or repeated outcome measurements. Functional regression models with longitudinal responses have been studied in the mean regression setting \citep{reiss2010fast,goldsmith2013corrected}, and extending these ideas to functional quantile regression would allow investigators to examine how time-varying physical activity patterns relate to trajectories of cognitive decline. Incorporating within-subject correlation structures into quantile-based functional models remains an open methodological challenge. Finally, while this work has focused on estimation and variable selection, valid inference after regularization remains challenging in high-dimensional settings. Standard inferential procedures do not directly apply following penalized selection. Future research will therefore investigate uncertainty quantification for VSFLQR, drawing on post-selection inference techniques \citep{lee2016exact, taylor2018post,belloni2019valid} and sample-splitting approaches \citep{wasserman2009high}.

\section*{Acknowledgement}
This work is supported by the SPARC Graduate Research grant from the Office of the Vice President for Research, University of South Carolina.

\section*{Supplementary Materials}
Web Appendices A, Tables 1-5, and Figures 1-6 referenced in Sections \ref{sec:method1}, \ref{sec:sim_stud}, and \ref{realdat} are available with this paper at the Biostatistics website. Software illustration of the proposed method is provided with this paper and will be made available online at Github.

\section*{Data Availability Statement}
The data supporting the findings of this study are publicly available at \cite{NHANES2011_2014}.

%\section*{Funding}

\singlespacing
\bibliographystyle{biorefs}
\bibliography{score}

@book{Ramsay05functionaldata,
    author = {J. Ramsay and B. Silverman},
    title = {Functional Data Analysis},
    Publisher={Springer-Verlag},
    address={New York},
    year = {2005}
}

@article{yue2026variable,
  title={Variable selection in functional linear Cox model},
  author={Yue, Yuanzhen and Self, Stella and Wu, Yichao and Zhang, Jiajia and Ghosal, Rahul},
  journal={Biometrics},
  volume={82},
  number={1},
  pages={ujag044},
  year={2026},
  publisher={Oxford University Press}
}

@article{henry2004verbal,
  title={Verbal fluency performance in dementia of the Alzheimer’s type: a meta-analysis},
  author={Henry, Julie D and Crawford, John R and Phillips, Louise H},
  journal={Neuropsychologia},
  volume={42},
  number={9},
  pages={1212--1222},
  year={2004},
  publisher={Elsevier}
}

@article{canning2004diagnostic,
  title={Diagnostic utility of abbreviated fluency measures in Alzheimer disease and vascular dementia},
  author={Canning, SJ Duff and Leach, Linda and Stuss, D and Ngo, L and Black, SE14981170},
  journal={Neurology},
  volume={62},
  number={4},
  pages={556--562},
  year={2004},
  publisher={Lippincott Williams \& Wilkins}
}

@article{zhou2024dietary,
  title={Dietary protein intake interacts with weak handgrip strength and cognitive impairment},
  author={Zhou, Lingling and Zhang, Cui},
  journal={Journal of Alzheimer’s Disease},
  volume={102},
  number={2},
  pages={359--369},
  year={2024},
  publisher={SAGE Publications Sage UK: London, England}
}

@book{ramsay2007applied,
  title={Applied Functional Data Analysis: Methods and Case Studies},
  author={Ramsay, J.O. and Silverman, B.W.},
  isbn={9780387224657},
  series={Springer Series in Statistics},
  url={https://books.google.com/books?id=WE3SzeVEvDkC},
  year={2007},
  publisher={Springer New York}
}

@article{besse2000autoregressive,
  title={Autoregressive forecasting of some functional climatic variations},
  author={Besse, Philippe C and Cardot, Herv{\'e} and Stephenson, David B},
  journal={Scandinavian Journal of Statistics},
  volume={27},
  number={4},
  pages={673--687},
  year={2000},
  publisher={Wiley Online Library}
}

@article{sherwood2025package,
    title = {rqPen: An R Package for Penalized Quantile Regression},
    author = {Ben Sherwood and Shaobo Li and Adam Maidman},
    journal = {The R Journal},
    year = {2025},
    volume = {17},
    number = {2},
    pages = {146-175},
    doi = {10.32614/RJ-2025-017},
  }

@article{li2022inference,
  title={Inference in functional linear quantile regression},
  author={Li, Meng and Wang, Kehui and Maity, Arnab and Staicu, Ana-Maria},
  journal={Journal of multivariate analysis},
  volume={190},
  pages={104985},
  year={2022},
  publisher={Elsevier}
}

@article{yao2017regularized,
  title={Regularized partially functional quantile regression},
  author={Yao, Fang and Sue-Chee, Shivon and Wang, Fan},
  journal={Journal of Multivariate Analysis},
  volume={156},
  pages={39--56},
  year={2017},
  publisher={Elsevier}
}

@article{hastie2015statistical,
  title={Statistical learning with sparsity},
  author={Hastie, Trevor and Tibshirani, Robert and Wainwright, Martin},
  journal={Monographs on statistics and applied probability},
  volume={143},
  number={143},
  pages={8},
  year={2015}
}

@article{cui2021additive,
  title={Additive functional Cox model},
  author={Cui, Erjia and Crainiceanu, Ciprian M and Leroux, Andrew},
  journal={Journal of Computational and Graphical Statistics},
  volume={30},
  number={3},
  pages={780--793},
  year={2021},
  publisher={Taylor \& Francis}
}

@article{ghosal2023functional,
  title={Functional proportional hazards mixture cure model with applications in cancer mortality in NHANES and post ICU recovery},
  author={Ghosal, Rahul and Matabuena, Marcos and Zhang, Jiajia},
  journal={Statistical Methods in Medical Research},
  volume={32},
  number={11},
  pages={2254--2269},
  year={2023},
  publisher={SAGE Publications Sage UK: London, England}
}

@article{tibshirani1996regression,
  title={Regression shrinkage and selection via the lasso},
  author={Tibshirani, Robert},
  journal={Journal of the Royal Statistical Society Series B: Statistical Methodology},
  volume={58},
  number={1},
  pages={267--288},
  year={1996},
  publisher={Oxford University Press}
}

@article{zhang2010nearly,
author={Zhang, Cun-Hui},
title = {{Nearly unbiased variable selection under minimax concave penalty}},
volume = {38},
journal = {The Annals of Statistics},
number = {2},
publisher = {Institute of Mathematical Statistics},
pages = {894 -- 942},
year = {2010},
doi = {10.1214/09-AOS729},
URL = {https://doi.org/10.1214/09-AOS729}
}

@article{chen2008extended,
  title={Extended Bayesian information criteria for model selection with large model spaces},
  author={Chen, Jiahua and Chen, Zehua},
  journal={Biometrika},
  volume={95},
  number={3},
  pages={759--771},
  year={2008},
  publisher={Oxford University Press}
}

@article{breheny2015group,
  title={Group descent algorithms for nonconvex penalized linear and logistic regression models with grouped predictors},
  author={Breheny, Patrick and Huang, Jian},
  journal={Statistics and computing},
  volume={25},
  pages={173--187},
  year={2015},
  publisher={Springer}
}

@article{ghosal2020variable,
  title={Variable selection in functional linear concurrent regression},
  author={Ghosal, Rahul and Maity, Arnab and Clark, Timothy and Longo, Stefano B},
  journal={Journal of the Royal Statistical Society Series C: Applied Statistics},
  volume={69},
  number={3},
  pages={565--587},
  year={2020},
  publisher={Oxford University Press}
}

@article{hosking1990moments,
  title={L-moments: analysis and estimation of distributions using linear combinations of order statistics},
  author={Hosking, Jonathan RM},
  journal={Journal of the Royal Statistical Society Series B: Statistical Methodology},
  volume={52},
  number={1},
  pages={105--124},
  year={1990},
  publisher={Oxford University Press}
}

@article{varma2017re,
  title={Re-evaluating the effect of age on physical activity over the lifespan},
  author={Varma, Vijay R and Dey, Debangan and Leroux, Andrew and Di, Junrui and Urbanek, Jacek and Xiao, Luo and Zipunnikov, Vadim},
  journal={Preventive medicine},
  volume={101},
  pages={102--108},
  year={2017},
  publisher={Elsevier}
}

@article{ghosal2022scalar,
  title={Scalar on time-by-distribution regression and its application for modelling associations between daily-living physical activity and cognitive functions in Alzheimer’s disease},
  author={Ghosal, Rahul and Varma, Vijay R and Volfson, Dmitri and Urbanek, Jacek and Hausdorff, Jeffrey M and Watts, Amber and Zipunnikov, Vadim},
  journal={Scientific reports},
  volume={12},
  number={1},
  pages={11558},
  year={2022},
  publisher={Nature Publishing Group UK London}
}

@article{goldsmith2016new,
  title={New insights into activity patterns in children, found using functional data analyses},
  author={Goldsmith, Jeff and Liu, Xinyue and Jacobson, Judith and Rundle, Andrew},
  journal={Medicine and science in sports and exercise},
  volume={48},
  number={9},
  pages={1723},
  year={2016},
  publisher={NIH Public Access}
}

@article{cho2024exploring,
  title={Exploring the association between daily distributional patterns of physical activity and cardiovascular mortality risk among older adults in NHANES 2003-2006},
  author={Cho, Sunwoo Emma and Saha, Enakshi and Matabuena, Marcos and Wei, Jingkai and Ghosal, Rahul},
  journal={Annals of Epidemiology},
  volume={99},
  pages={24--31},
  year={2024},
  publisher={Elsevier}
}

@article{koenker1978regression,
 ISSN = {00129682, 14680262},
 URL = {http://www.jstor.org/stable/1913643},
 author = {Koenker, Roger and Bassett Jr, Gilbert},
 journal = {Econometrica},
 number = {1},
 pages = {33--50},
 publisher = {[Wiley, Econometric Society]},
 title = {Regression Quantiles},
 urldate = {2026-03-23},
 volume = {46},
 year = {1978}
}

@book{koenker2005quantile,
    place={Cambridge},
    series={Econometric Society Monographs},
    title={Quantile Regression},
    publisher={Cambridge University Press},
    author={Koenker, Roger},
    year={2005},
    collection={Econometric Society Monographs}
}

@book{davino2013quantile,
  title={Quantile regression: theory and applications},
  author={Davino, Cristina and Furno, Marilena and Vistocco, Domenico},
  volume={988},
  year={2013},
  publisher={John Wiley \& Sons}
}

@article{koenker2017quantile,
  title={Quantile regression: 40 years on},
  author={Koenker, Roger},
  journal={Annual review of economics},
  volume={9},
  number={1},
  pages={155--176},
  year={2017},
  publisher={Annual Reviews}
}

@article{koenker2001quantile,
  title={Quantile regression},
  author={Koenker, Roger and Hallock, Kevin F},
  journal={Journal of economic perspectives},
  volume={15},
  number={4},
  pages={143--156},
  year={2001},
  publisher={American Economic Association}
}

@article{takeuchi2006nonparametric,
author={Takeuchi, Ichiro and Le, Quoc V and Sears, Timothy D and Smola, Alexander J and Williams, Chris},
year = {2006},
month = {12},
pages = {1231-1264},
title = {Nonparametric Quantile Estimation},
volume = {7},
journal = {Journal of Machine Learning Research}
}

@article{yu2001bayesian,
  title={Bayesian quantile regression},
  author={Yu, Keming and Moyeed, Rana A},
  journal={Statistics \& Probability Letters},
  volume={54},
  number={4},
  pages={437--447},
  year={2001},
  publisher={Elsevier}
}

@article{mazumder2011sparsenet,
  title={Sparsenet: Coordinate descent with nonconvex penalties},
  author={Mazumder, Rahul and Friedman, Jerome H and Hastie, Trevor},
  journal={Journal of the American Statistical Association},
  volume={106},
  number={495},
  pages={1125--1138},
  year={2011},
  publisher={Taylor \& Francis}
}

@article{yi2017semismooth,
  title={Semismooth newton coordinate descent algorithm for elastic-net penalized huber loss regression and quantile regression},
  author={Yi, Congrui and Huang, Jian},
  journal={Journal of Computational and Graphical Statistics},
  volume={26},
  number={3},
  pages={547--557},
  year={2017},
  publisher={Taylor \& Francis}
}

@article{yao2005functional,
  title={Functional data analysis for sparse longitudinal data},
  author={Yao, Fang and M{\"u}ller, Hans-Georg and Wang, Jane-Ling},
  journal={Journal of the American statistical association},
  volume={100},
  number={470},
  pages={577--590},
  year={2005},
  publisher={Taylor \& Francis}
}

@incollection{huber1992robust,
  title={Robust estimation of a location parameter},
  author={Huber, Peter J},
  booktitle={Breakthroughs in statistics: Methodology and distribution},
  pages={492--518},
  year={1992},
  publisher={Springer}
}

@article{sherwood2022quantile,
  title={Quantile regression feature selection and estimation with grouped variables using Huber approximation},
  author={Sherwood, Ben and Li, Shaobo},
  journal={Statistics and Computing},
  volume={32},
  number={5},
  pages={75},
  year={2022},
  publisher={Springer}
}

@article{yuan2006model,
  title={Model selection and estimation in regression with grouped variables},
  author={Yuan, Ming and Lin, Yi},
  journal={Journal of the Royal Statistical Society Series B: Statistical Methodology},
  volume={68},
  number={1},
  pages={49--67},
  year={2006},
  publisher={Oxford University Press}
}

@article{kato2012estimation,
 ISSN = {00905364, 21688966},
 URL = {http://www.jstor.org/stable/41806568},
 author = {Kato, Kengo},
 journal = {The Annals of Statistics},
 number = {6},
 pages = {3108--3136},
 publisher = {Institute of Mathematical Statistics},
 title = {ESTIMATION IN FUNCTIONAL LINEAR QUANTILE REGRESSION},
 urldate = {2026-03-23},
 volume = {40},
 year = {2012}
}

@article{chen2016variable,
  title={Variable selection in function-on-scalar regression},
  author={Chen, Yakuan and Goldsmith, Jeff and Ogden, R Todd},
  journal={Stat},
  volume={5},
  number={1},
  pages={88--101},
  year={2016},
  publisher={Wiley Online Library}
}

@article{ghosal2023variable,
  title={Variable selection in nonlinear function-on-scalar regression},
  author={Ghosal, Rahul and Maity, Arnab},
  journal={Biometrics},
  volume={79},
  number={1},
  pages={292--303},
  year={2023},
  publisher={Wiley Online Library}
}

@article{yau2025physical,
  title={Physical activity as a modifiable risk factor in preclinical Alzheimer’s disease},
  author={Yau, Wai-Ying Wendy and Kirn, Dylan R and Rabin, Jennifer S and Properzi, Michael J and Schultz, Aaron P and Shirzadi, Zahra and Palmgren, Kailee and Matos, Paola and Maa, Courtney and Pruzin, Jeremy J and others},
  journal={Nature Medicine},
  pages = {4075-4083},
  volume = {31},
  year={2025},
  publisher={Nature Publishing Group US New York}
}

@article{campbell2023estimating,
  title={Estimating the effect of physical activity on cognitive function within the UK Biobank cohort},
  author={Campbell, Thomas and Cullen, Breda},
  journal={International Journal of Epidemiology},
  volume={52},
  number={5},
  pages={1592--1611},
  year={2023},
  publisher={Oxford University Press}
}

@article{quinlan2023physical,
  title={Physical activity and cognitive function in middle-aged adults: a cross-sectional analysis of the PATH through life study},
  author={Quinlan, Clare and Rattray, Ben and Pryor, Disa and Northey, Joseph M and Cherbuin, Nicolas},
  journal={Frontiers in Psychology},
  volume={14},
  pages={1022868},
  year={2023},
  publisher={Frontiers Media SA}
}

@article{ghosal2023distributional,
  title={Distributional data analysis via quantile functions and its application to modeling digital biomarkers of gait in Alzheimer’s disease},
  author={Ghosal, Rahul and Varma, Vijay R and Volfson, Dmitri and Hillel, Inbar and Urbanek, Jacek and Hausdorff, Jeffrey M and Watts, Amber and Zipunnikov, Vadim},
  journal={Biostatistics},
  volume={24},
  number={3},
  pages={539--561},
  year={2023},
  publisher={Oxford University Press}
}

@article{xiao2015quantifying,
  title={Quantifying the lifetime circadian rhythm of physical activity: a covariate-dependent functional approach},
  author={Xiao, Luo and Huang, Lei and Schrack, Jennifer A and Ferrucci, Luigi and Zipunnikov, Vadim and Crainiceanu, Ciprian M},
  journal={Biostatistics},
  volume={16},
  number={2},
  pages={352--367},
  year={2015},
  publisher={Oxford University Press}
}

@article{vogel1993moment,
  title={L moment diagrams should replace product moment diagrams},
  author={Vogel, Richard M and Fennessey, Neil M},
  journal={Water resources research},
  volume={29},
  number={6},
  pages={1745--1752},
  year={1993},
  publisher={Wiley Online Library}
}

@article{wu2009variable,
 ISSN = {10170405, 19968507},
 URL = {http://www.jstor.org/stable/24308857},
 author = {Wu, Yichao and Liu, Yufeng},
 journal = {Statistica Sinica},
 number = {2},
 pages = {801--817},
 publisher = {Institute of Statistical Science, Academia Sinica},
 title = {VARIABLE SELECTION IN QUANTILE REGRESSION},
 urldate = {2026-03-23},
 volume = {19},
 year = {2009}
}

@article{su2021elastic,
  title={Elastic net penalized quantile regression model},
  author={Su, Meihong and Wang, Wenjian},
  journal={Journal of Computational and Applied Mathematics},
  volume={392},
  pages={113462},
  year={2021},
  publisher={Elsevier}
}

@article{dai2023high,
  title={High-dimensional variable selection for quantile regression based on variational bayesian method},
  author={Dai, Dengluan and Tang, Anmin and Ye, Jinli},
  journal={Mathematics},
  volume={11},
  number={10},
  pages={2232},
  year={2023},
  publisher={MDPI}
}

@article{tang2014partial,
  title={Partial functional linear quantile regression},
  author={Tang, QingGuo and Cheng, LongSheng},
  journal={Science China Mathematics},
  volume={57},
  number={12},
  pages={2589--2608},
  year={2014},
  publisher={Springer}
}

@article{ma2019quantile,
  title={Quantile regression for functional partially linear model in ultra-high dimensions},
  author={Ma, Haiqiang and Li, Ting and Zhu, Hongtu and Zhu, Zhongyi},
  journal={Computational Statistics \& Data Analysis},
  volume={129},
  pages={135--147},
  year={2019},
  publisher={Elsevier}
}

@book{crainiceanu2024functional,
  title={Functional data analysis with R},
  author={Crainiceanu, Ciprian M and Goldsmith, Jeff and Leroux, Andrew and Cui, Erjia},
  year={2024},
  publisher={CRC Press}
}

@article{ratcliffe2002functional,
  title={Functional data analysis with application to periodically stimulated foetal heart rate data. II: Functional logistic regression},
  author={Ratcliffe, Sarah J and Heller, Gillian Z and Leader, Leo R},
  journal={Statistics in medicine},
  volume={21},
  number={8},
  pages={1115--1127},
  year={2002},
  publisher={Wiley Online Library}
}

@article{diller2006heart,
  title={Heart rate response during exercise predicts survival in adults with congenital heart disease},
  author={Diller, Gerhard-Paul and Dimopoulos, Konstantinos and Okonko, Darlington and Uebing, Anselm and Broberg, Craig S and Babu-Narayan, Sonya and Bayne, Stephanie and Poole-Wilson, Philip A and Sutton, Richard and Francis, Darrel P and others},
  journal={Journal of the American College of Cardiology},
  volume={48},
  number={6},
  pages={1250--1256},
  year={2006},
  publisher={American College of Cardiology Foundation Washington, DC}
}

@article{cui2022fast,
  title={Fast univariate inference for longitudinal functional models},
  author={Cui, Erjia and Leroux, Andrew and Smirnova, Ekaterina and Crainiceanu, Ciprian M},
  journal={Journal of Computational and Graphical Statistics},
  volume={31},
  number={1},
  pages={219--230},
  year={2022},
  publisher={Taylor \& Francis}
}

@article{tian2010functional,
  title={Functional data analysis in brain imaging studies},
  author={Tian, Tian Siva},
  journal={Frontiers in psychology},
  volume={1},
  pages={35},
  year={2010},
  publisher={Frontiers Research Foundation}
}

@article{du2018estimation,
  title={Estimation and variable selection for partially functional linear models},
  author={Du, Jiang and Xu, Dengke and Cao, Ruiyuan},
  journal={Journal of the Korean Statistical Society},
  volume={47},
  number={4},
  pages={436--449},
  year={2018},
  publisher={Elsevier}
}

@article{wang2023smoothed,
  title={Smoothed quantile regression for partially functional linear models in high dimensions},
  author={Wang, Zhihao and Bai, Yongxin and H{\"a}rdle, Wolfgang K and Tian, Maozai},
  journal={Biometrical Journal},
  volume={65},
  number={7},
  pages={2200060},
  year={2023},
  publisher={Wiley Online Library}
}

@article{donahue2025activity,
  title={Activity variability: A novel physical activity metric and its association with cognitive impairment},
  author={Donahue, Patrick T and Schrack, Jennifer A and Thrul, Johannes and Carlson, Michelle C},
  journal={Alzheimer's \& Dementia: Translational Research \& Clinical Interventions},
  volume={11},
  number={2},
  pages={e70079},
  year={2025},
  publisher={Wiley Online Library}
}

@article{john2019open,
  title={An open-source monitor-independent movement summary for accelerometer data processing},
  author={John, Dinesh and Tang, Qu and Albinali, Fahd and Intille, Stephen},
  journal={Journal for the measurement of physical behaviour},
  volume={2},
  number={4},
  pages={268--281},
  year={2019},
  publisher={Human Kinetics}
}

@article{anderson2007cognitive,
  title={Cognitive health: an emerging public health issue},
  author={Anderson, Lynda A and McConnell, Stephen R},
  journal={Alzheimers Dement},
  volume={3},
  number={2 Suppl},
  pages={S70--S73},
  year={2007}
}

@article{matsui2026decline,
  author={Matsui, Yosuke and Fujisawa, Chisato and Minakami, Miki and Yamada, Yosuke and Watanabe, Kazuhisa and Nakashima, Hirotaka and Komiya, Hitoshi and Umegaki, Hiroyuki},
  title     = {The decline in cognitive function with age and its changes over time in cognitively normal older adults},
  journal   = {European Geriatric Medicine},
  year      = {2026},
  volume    = {17},
  number    = {1},
  pages     = {299--308},
  doi       = {10.1007/s41999-025-01377-8}
}

@incollection{yang2023cognitive,
  title={Cognitive decline associated with aging},
  author={Yang, Yiru and Wang, Dandan and Hou, Wenjie and Li, He},
  booktitle={Cognitive aging and brain health},
  pages={25--46},
  year={2023},
  publisher={Springer}
}

@article{mani2013poverty,
  title={Poverty impedes cognitive function},
  author={Mani, Anandi and Mullainathan, Sendhil and Shafir, Eldar and Zhao, Jiaying},
  journal={science},
  volume={341},
  number={6149},
  pages={976--980},
  year={2013},
  publisher={American Association for the Advancement of Science}
}

@article{krueger2025lifetime,
  title={Lifetime socioeconomic status, cognitive decline, and brain characteristics},
  author={Krueger, Kristin R and Desai, Pankaja and Beck, Todd and Barnes, Lisa L and Bond, Jerenda and DeCarli, Charles and Aggarwal, Neelum T and Evans, Denis A and Rajan, Kumar B},
  journal={JAMA network open},
  volume={8},
  number={2},
  pages={e2461208},
  year={2025}
}

@article{clouston2020education,
  title={Education and cognitive decline: an integrative analysis of global longitudinal studies of cognitive aging},
  author={Clouston, Sean AP and Smith, Dylan M and Mukherjee, Soumyadeep and Zhang, Yun and Hou, Wei and Link, Bruce G and Richards, Marcus},
  journal={The Journals of Gerontology: Series B},
  volume={75},
  number={7},
  pages={e151--e160},
  year={2020},
  publisher={Oxford University Press US}
}

@article{seyedsalehi2023educational,
  title={Educational attainment, structural brain reserve and Alzheimer’s disease: a Mendelian randomization analysis},
  author={Seyedsalehi, Aida and Warrier, Varun and Bethlehem, Richard AI and Perry, Benjamin I and Burgess, Stephen and Murray, Graham K},
  journal={Brain},
  volume={146},
  number={5},
  pages={2059--2074},
  year={2023},
  publisher={Oxford University Press US}
}

@article{wu2007controlling,
  title={Controlling variable selection by the addition of pseudovariables},
  author={Wu, Yujun and Boos, Dennis D and Stefanski, Leonard A},
  journal={Journal of the American Statistical Association},
  volume={102},
  number={477},
  pages={235--243},
  year={2007},
  publisher={Taylor \& Francis}
}

@article{feng2023associations,
  title={Associations of timing of physical activity with all-cause and cause-specific mortality in a prospective cohort study},
  author={Feng, Hongliang and Yang, Lulu and Liang, Yannis Yan and Ai, Sizhi and Liu, Yaping and Liu, Yue and Jin, Xinyi and Lei, Binbin and Wang, Jing and Zheng, Nana and others},
  journal={Nature communications},
  volume={14},
  number={1},
  pages={930},
  year={2023},
  publisher={Nature Publishing Group UK London}
}

@article{horowitz2005nonparametric,
  title={Nonparametric estimation of an additive quantile regression model},
  author={Horowitz, Joel L and Lee, Sokbae},
  journal={Journal of the American Statistical Association},
  volume={100},
  number={472},
  pages={1238--1249},
  year={2005},
  publisher={Taylor \& Francis}
}

@article{waldmann2013bayesian,
  title={Bayesian semiparametric additive quantile regression},
  author={Waldmann, Elisabeth and Kneib, Thomas and Yue, Yu Ryan and Lang, Stefan and Flexeder, Claudia},
  journal={Statistical Modelling},
  volume={13},
  number={3},
  pages={223--252},
  year={2013},
  publisher={SAGE Publications Sage India: New Delhi, India}
}

@article{zhang2021functional,
  title={Functional additive quantile regression},
  author={Zhang, Yingying and Lian, Heng and Li, Guodong and Zhu, Zhongyi},
  journal={Statistica Sinica},
  volume={31},
  number={3},
  pages={1331--1351},
  year={2021},
  publisher={JSTOR}
}

@article{lee2016exact,
author = {Lee, Jason D and Sun, Dennis L and Sun, Yuekai and Taylor, Jonathan E},
title = {{Exact post-selection inference, with application to the lasso}},
volume = {44},
journal = {The Annals of Statistics},
number = {3},
publisher = {Institute of Mathematical Statistics},
pages = {907 -- 927},
year = {2016},
doi = {10.1214/15-AOS1371},
URL = {https://doi.org/10.1214/15-AOS1371}
}

@article{taylor2018post,
  title={Post-selection inference for-penalized likelihood models},
  author={Taylor, Jonathan and Tibshirani, Robert},
  journal={Canadian Journal of Statistics},
  volume={46},
  number={1},
  pages={41--61},
  year={2018},
  publisher={Wiley Online Library}
}

@article{wasserman2009high,
  title={High dimensional variable selection},
  author={Wasserman, Larry and Roeder, Kathryn},
  journal={Annals of statistics},
  volume={37},
  number={5A},
  pages={2178},
  year={2009}
}

@article{zou2008composite,
 ISSN = {00905364},
 URL = {http://www.jstor.org/stable/25464661},
 author = {Zou, Hui and Yuan, Ming},
 journal = {The Annals of Statistics},
 number = {3},
 pages = {1108--1126},
 publisher = {Institute of Mathematical Statistics},
 title = {Composite Quantile Regression and the Oracle Model Selection Theory},
 urldate = {2026-03-23},
 volume = {36},
 year = {2008}
}

@article{belloni2019valid,
  title={Valid post-selection inference in high-dimensional approximately sparse quantile regression models},
  author={Belloni, Alexandre and Chernozhukov, Victor and Kato, Kengo},
  journal={Journal of the American Statistical Association},
  volume={114},
  number={526},
  pages={749--758},
  year={2019},
  publisher={Taylor \& Francis}
}

@article{goldsmith2013corrected,
  title={Corrected confidence bands for functional data using principal components},
  author={Goldsmith, Jeff and Greven, Sonja and Crainiceanu, CIPRIAN},
  journal={Biometrics},
  volume={69},
  number={1},
  pages={41--51},
  year={2013},
  publisher={Oxford University Press}
}

@article{reiss2010fast,
journal={The International Journal of Biostatistics},
author={Reiss Philip T. and Huang Lei and Mennes Maarten},
title={Fast Function-on-Scalar Regression with Penalized Basis Expansions},
year={2010},
month={August},
pages={1-30},
volume={6},
number={1},
doi={10.2202/1557-4679.1246},
url={https://ideas.repec.org/a/bpj/ijbist/v6y2010i1n28.html},
}

@misc{NHANES2011_2014,
  author       = {{National Health and Nutrition Examination Survey (NHANES)}},
  year         = {2014},
  howpublished = {\url{https://wwwn.cdc.gov/nchs/nhanes/default.aspx}},
  note         = {NHANES 2011--2012 and 2013--2014 public-use data files; accessed June 3, 2025}
}

@misc{VSFLQR,
  author       = {Yue, Yuanzhen and Self, Stella and Wu, Yichao and Zhang, Jiajia and Ghosal, Rahul},
  year         = {2026},
  howpublished = {\url{https://github.com/DiamondMine97}},
  note         = {Software implementation of Variable Selection in Functional Linear Quantile Regression (VSFLQR).; accessed February 12, 2026}
}

 \newpage
%\begin{figure}[ht]
%\centering
%\includegraphics[width=0.7\linewidth,height=0.45\linewidth]{Figures/Motivating_plot.png}
%\caption{Within-day temporal pattern of mood ratings ($1$ = Most Happy to $7$ = Most Sad, $4$ = Neutral) over $7$ days for four different subjects in National Institute of Mental Health Family Study of Mood Disorders Subtypes.}
%\label{fig:motivation}
%\end{figure}

\end{document}